\documentclass[
superscriptaddress,
amsmath,amssymb,
aps,
prb,
twocolumn, floatfix, reprint,
longbibliography
]{revtex4-2}

\usepackage{graphicx}
\usepackage{dcolumn}
\usepackage{bm}

\usepackage{tensor}
\usepackage{accents}
\usepackage[shortlabels]{enumitem}
\usepackage[caption=false]{subfig} 
\usepackage{xcolor}

\usepackage[colorlinks,urlcolor=blue,linkcolor=blue,citecolor=blue,breaklinks=True]{hyperref} 
\usepackage[capitalize,noabbrev]{cleveref} 

\newcommand{\ket}[1]{\left\lvert #1 \right\rangle} 
\newcommand{\outerprod}[2]{{\left\vert #1 \vphantom{#2} \right\rangle} {\left\langle #2 \vphantom{#1} \right\vert}} 

\begin{document}

\preprint{APS/123-QED}

\title{False vacuum decay and nucleation dynamics in neutral atom systems}

\author{Siva~Darbha}
 \email{sdarbha@lbl.gov}
 \email{siva.darbha@berkeley.edu}
 \affiliation{Applied Mathematics and Computational Research Division, Lawrence Berkeley National Laboratory, Berkeley, CA 94720, USA}
 \affiliation{National Energy Research Scientific Computing Center, Lawrence Berkeley National Laboratory, Berkeley, CA 94720, USA}

\author{Milan~Kornja\v{c}a}
 \affiliation{QuEra Computing Inc., 1284 Soldiers Field Road, Boston, MA 02135, USA}

\author{Fangli~Liu}
 \affiliation{QuEra Computing Inc., 1284 Soldiers Field Road, Boston, MA 02135, USA}

\author{Jan~Balewski}
 \affiliation{National Energy Research Scientific Computing Center, Lawrence Berkeley National Laboratory, Berkeley, CA 94720, USA}

\author{Mark~R.~Hirsbrunner}
 \affiliation{National Energy Research Scientific Computing Center, Lawrence Berkeley National Laboratory, Berkeley, CA 94720, USA}
 \affiliation{Department of Physics and Institute for Condensed Matter Theory, University of Illinois at Urbana-Champaign, Urbana, IL 61801, USA}

\author{Pedro~L.~S.~Lopes}
 \affiliation{QuEra Computing Inc., 1284 Soldiers Field Road, Boston, MA 02135, USA}

\author{Sheng-Tao~Wang}
 \affiliation{QuEra Computing Inc., 1284 Soldiers Field Road, Boston, MA 02135, USA}

\author{Roel~Van~Beeumen}
 \affiliation{Applied Mathematics and Computational Research Division, Lawrence Berkeley National Laboratory, Berkeley, CA 94720, USA}

\author{Daan~Camps}
 \affiliation{National Energy Research Scientific Computing Center, Lawrence Berkeley National Laboratory, Berkeley, CA 94720, USA}

\author{Katherine~Klymko}
 \affiliation{National Energy Research Scientific Computing Center, Lawrence Berkeley National Laboratory, Berkeley, CA 94720, USA}


\begin{abstract}
Metastable states of quantum many-body systems with confinement offer a means to simulate false vacuum phenomenology, including non-equilibrium dynamical processes like decay by nucleation, in truncated limits. Recent work has examined the decay process in 1D ferromagnetic Ising spins and superfluids. In this paper, we study nucleation dynamics in 1D antiferromagnetic neutral atom chains with Rydberg interactions, using both numerical simulations and analytic modeling. We apply a staggered local detuning field to generate the metastable and ground states. Our efforts focus on two dynamical regimes: decay and annealing. In the first, we corroborate the phenomenological decay rate scaling and determine the associated parameter range for the decay process; in the second, we uncover and elucidate a procedure to anneal the metastable state from the initial to the final system, with intermediate nucleation events. We further propose experimental protocols to prepare the required states and perform quenches on near-term neutral atom quantum simulators, examining the experimental feasibility of our proposed setup and parameter regime.
\end{abstract}

\maketitle

\section{Introduction}
\label{sec:intro}

Metastable states in quantum systems lie above the true ground state and can decay by tunneling via nucleation~\cite{Coleman:1977a,Coleman:1977b}, thereby transforming under a first-order phase transition. They were examined in pioneering studies in quantum field theory as false vacua~\cite{Kobzarev:1974cp,Stone:1976,Stone:1977,Coleman:1977a,Coleman:1977b,Callan:1977,Voloshin:1985id}, after earlier analysis in statistical physics of similar phenomena~\cite{Langer:1967,Lifshitz:1972}. Metastable states and their decay properties have also been explored in strongly-correlated quantum systems, including ultracold atoms~\cite{Hruby:2018,Song:2022,Zenesini:2024} and magnetic materials~\cite{Rutkevich:1999,Lagnese:2021,Lagnese:2023}. In addition, more general metastable states in quasi-classical systems can also undergo phase transitions, including macromolecules such as polymers~\cite{Cheng:1998,Keller:1998} and proteins~\cite{Thirumalai:2011,Ghosh:2020}, possibly influenced by similar effects.

In quantum field theory, false vacuum decay was originally studied for scalar fields without gravitational effects, for its applications in early-universe cosmology~\cite{Kobzarev:1974cp,Coleman:1977a,Coleman:1977b,Callan:1977} and fundamental interactions~\cite{Stone:1976,Stone:1977,Voloshin:1985id}. The decay occurs by a tunneling process in which true vacuum bubbles are nucleated by quantum fluctuations in the false vacuum background, and bubbles larger than a critical size can rapidly expand throughout the state~\cite{Coleman:1977a,Coleman:1977b}. The decay rate, that is, the tunneling probability per unit time, has been derived analytically under simplifying assumptions~\cite{Kobzarev:1974cp,Stone:1976,Stone:1977,Coleman:1977a,Coleman:1977b,Callan:1977,Voloshin:1985id}. In $1+1$ spacetime dimensions, for certain potentials and in the ``thin-wall'' approximation (elaborated in \Cref{subsubsec:decay:rate:qft}), it has the characteristic scaling $\gamma \propto \mathcal{E} e^{-\lambda \mathcal{E}^{-1}}$, where $\mathcal{E}$ is the energy density difference between the false and true vacua, and $\lambda$ collects the remaining dependencies~\cite{Stone:1977,Voloshin:1985id}. For small $\mathcal{E}$, the decay rate becomes exponentially small, a feature of the metastability. Recent numerical simulations with quantum quenches have corroborated this scaling for $(1+1)$-dimensional $\phi^4$~\cite{SzaszSchagrin:2022} and Ising~\cite{Lencses:2022} field theories, with additional fitting parameters in the prefactor. In cosmology, false vacuum studies have been extended to incorporate additional physics, namely general relativity~\cite{Coleman:1980,Guth:1983} and finite temperature~\cite{Linde:1981,Linde:1983}. Concomitant efforts also examined the impact of the possible nucleation, expansion, and collision of true vacuum bubbles on the evolution of the Universe and its field content~\cite{Guth:1981,Turner:1982,Hawking:1982,Guth:1983}. For such a cosmological phase transition~\cite{Mazumdar:2019,Hindmarsh:2021}, observable signatures could manifest in various different channels, including the Higgs mass~\cite{Markkanen:2018} and gravitational wave signals~\cite{Athron:2024}.

Metastable states and their decay are also important features of strongly-correlated quantum matter and materials. Ultracold atoms are effective systems to study these states due to their controllability, and a number of experiments employing them have been proposed~\cite{Opanchuk:2013,Fialko:2015,Fialko:2017,Braden:2018,Billam:2019,Braden:2019,Billam:2020,Ng:2021,Billam:2021,Jenkins:2023,Jenkins:2024} and implemented~\cite{Hruby:2018,Song:2022,Zenesini:2024}. Notably, the authors of Ref.~\cite{Zenesini:2024} recently performed an experiment that used ferromagnetic superfluids, formed from atomic Bose-Einstein condensates, and an adiabatic protocol to observe signatures of the decay process driven by thermal and quantum fluctuations, realizing the first experimental test of the decay rate scaling. Quantum spin chains also offer many advantages for studying metastability~\cite{Rutkevich:1999,Lagnese:2021,Lagnese:2023}, particularly due to advances in the use of quantum quenches and adiabatic ramps~\cite{Kormos:2017,Liu:2019,Tan:2021}. A number of exegetic results have been obtained from these systems. The decay rate has been derived analytically in the perturbative limit~\cite{Rutkevich:1999}, providing the spin analogue of the expression from quantum field theory. Numerical simulations with quantum quenches have recently exhibited and corroborated the exponential scaling~\cite{Lagnese:2021}. More general nucleation dynamics have been observed outside of the decay regime. Numerical investigations have examined linear ramps across first-order phase transitions and quantified the nucleation points for sequential domain sizes~\cite{Sinha:2021}. Real-time simulations have studied the collapse of bubbles~\cite{Milsted:2022}. Scalar field theories have also been encoded in spin systems, with metastable states detected through simulated annealing~\cite{Abel:2021}. 

The false and true vacuum states of quantum fields differ from the metastable and ground states of truncated quantum many-body systems, in that bosonic quantum fields include all field modes whereas finite-dimensional quantum systems have comparatively small Hilbert spaces. However, with these qualifications, we retain the former terms in the latter context due to the conceptual concordance between the two phenomena and the established terminology conventions in the literature (e.g.~\cite{Lagnese:2021,Lagnese:2023,Milsted:2022,Zenesini:2024}). Importantly, the decay process also occurs through nonperturbative bubble nucleation and the decay rate scaling exhibits a similar form. Experiments with these systems can investigate the reduction of field theoretical predictions to truncated models, and potentially motivate further theoretical investigations.

Neutral atoms with Rydberg excitations have emerged as effective systems to study quantum many-body dynamics~\cite{Browaeys:2020}, due to experimental innovations allowing efficient entanglement generation~\cite{Jaksch:2000,Lukin:2001,Gaetan:2009,Urban:2009}, control of individual atoms~\cite{Labuhn:2014,Kaufman:2014,Endres:2016}, and arrangement of large atomic arrays~\cite{Nogrette:2014,Barredo:2016,Labuhn:2016,Scholl:2021,Ebadi:2021}. These systems have been central in the discovery of quantum many-body scars~\cite{Bernien:2017,Turner:2018,Bluvstein:2021}; the verification of the quantum Kibble-Zurek mechanism for second-order phase transitions~\cite{Keesling:2019,Ebadi:2021}; the characterization of ground state phase diagrams in one and two dimensions~\cite{Samajdar:2020,Samajdar:2021,Ohler:2022,Chen:2023}, along with the realization of exotic phases~\cite{Lienhard:2020,Zhang:2024}; the exploration of topological phases~\cite{deLeseleuc:2019}, including topological spin liquids~\cite{Verresen:2021,Semeghini:2021,Ohler:2023,Kornjaca:2023}; and the simulation of lattice gauge theories~\cite{Surace:2020,Surace:2021}. In addition, these systems have been adapted for both digital quantum logic~\cite{Isenhower:2010,Saffman:2010} and analog quantum simulation~\cite{Henriet:2020,aquila2023quera}. 

In this paper, we study false vacuum decay and nucleation in a 1D neutral atom chain with Rydberg interactions. We use numerical methods and analytic modeling. Our work undertakes an extensive investigation of this antiferromagnetic system, and inherently includes finite-size effects and long-range interactions. We explicitly examine setups that are accessible to near-term experiments and propose experimental protocols for near-term quantum simulators. Our efforts focus on two regimes that yield nucleation dynamics: decay and annealing. In the decay regime, where the false vacuum decays by tunneling via nucleation, we determine the parameter range to observe the tunneling process, and quantify the decay rate scaling with the confinement parameter and the gap structure. In the annealing regime, where we tune the energy landscape to reverse the false and true vacuum states, we uncover a procedure to adiabatically transfer the false vacuum from the initial to the final system, with quantized nucleation events at condensation and decondensation points.

We further study long-lived oscillations of false and true vacua in a concurrent paper~\cite{Darbha:2024b}. That process arises in a more general parameter regime, and can be used to distinguish between the two vacuum states.

The paper is organized as follows. We describe our neutral atom model in \Cref{sec:model}. We examine the regime of false vacuum decay in \Cref{sec:decay} and annealing in \Cref{sec:annealing}. In \Cref{sec:experiments}, we highlight the experimental feasibility of our setups and propose experimental protocols for near-term neutral atom quantum hardware. \Cref{sec:conclusion} concludes and describes future prospects.
\section{Model}
\label{sec:model}

\begin{figure}
\subfloat[Decay regime]{\label{fig:overview_decay}\includegraphics[width=0.48\textwidth]{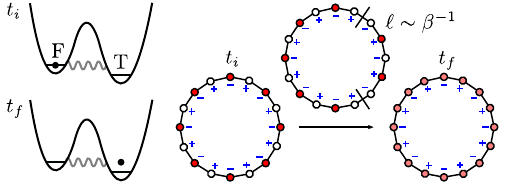}}\\
\subfloat[Annealing regime]{\label{fig:overview_annealing}\includegraphics[width=0.48\textwidth]{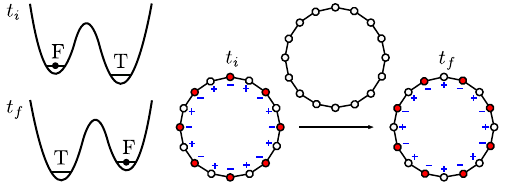}}
\caption{A schematic overview of our system and dynamical regimes. We consider a 1D neutral atom chain arranged as a ring. We apply the staggered local detuning field $\Delta_{\mathrm{loc},j}$ in \Cref{eq:local_detuning} to produce an asymmetric potential well with false and true vacuum states; the detuning sign at each site is shown by the blue $+/-$ pattern. The state is initialized to the false vacuum state, dominated by the $Z_2$ product state $\ket{10 \hdots 10}$, shown by the alternating color pattern where red indicates a Rydberg state and white indicates a ground state. (a) The decay regime. A constant positive local detuning field is applied. The state then tunnels through the potential barrier by the nucleation of domains above a critical size $\ell \sim \beta^{-1}$, and at finite temperature approaches a late-time thermal state, indicated by the blended red/white color pattern. (b) The annealing regime. A linear ramp of the local detuning is applied from a large positive to a large negative value, which smoothly tunes the potential well from one asymmetry to another. The state is then annealed from the false vacuum of the initial to the final system, passing through the all-zero state via nucleation.}
\label{fig:overview}
\end{figure}

We explore nucleation dynamics in the regimes of decay and annealing. \Cref{fig:overview} schematically shows the dynamics in these two regimes, which will be motivated and described in Sections \ref{sec:decay} and \ref{sec:annealing}. In our numerical studies, we initialize the state near the false vacuum, evolve it under the Hamiltonian, and quantify it using various observables. In this section, we describe our Hamiltonian, input parameters, confinement behavior, and relevant observables. 

We study a 1D neutral atom chain, as shown in \Cref{fig:overview}. It has an even number of sites $n_s$, where we choose $n_s = 16$ as a specific example to obtain reasonable computational wall times and enable a broad parameter study; a constant atom separation $a$; and periodic boundary conditions (PBCs) with the atoms arranged in a circle. Each atom $j = 1, \hdots, n_s$ is treated as a two-level system with a ground state $\ket{g_j} \equiv \ket{0_j}$ and a Rydberg state $\ket{r_j} \equiv \ket{1_j}$. 

The system is described by the Rydberg Hamiltonian
\begin{equation}
H = \frac{\Omega}{2} \sum_j \hat{\sigma}_{x,j} - \sum_j \Delta_j \hat{n}_j + \sum_{j<k} V_{jk} \hat{n}_j \hat{n}_k \, ,
\label{eq:hamiltonian}
\end{equation}
where $\hat{\sigma}_{x,j} = \outerprod{g_j}{r_j} + \outerprod{r_j}{g_j}$, $\hat{n}_j = \outerprod{r_j}{r_j}$, $\Omega$ is the Rabi frequency, $\Delta_j$ is the detuning field, and $V_{jk} = C_6 / r_{jk}^6$ is the two-site potential arising from the van der Waals interaction between Rydberg states, with $C_6$ a constant that depends on the atom species and $r_{jk}$ the separation between sites $j$ and $k$. The most important potentials are $V_1 = C_6 / r_{i,i+1}^6$ for nearest-neighbor and $V_2 = C_6 / r_{i,i+2}^6$ for next-nearest-neighbor interactions. The detuning field can be decomposed into global and local pieces,
\begin{equation}
\Delta_j = \Delta_\mathrm{glob} + \Delta_{\mathrm{loc},j} \, .
\end{equation}
The competition between the Rydberg coupling and the Rabi frequency produces the Rydberg blockade mechanism for atoms separated by roughly less than a blockade radius $R_b = (C_6 / \Omega)^{1/6}$, since joint excitations to the Rydberg state are suppressed by the large energy penalty from the $V_{jk} \propto 1/r_{jk}^6$ scaling~\cite{Jaksch:2000,Lukin:2001}.

We use a constant Rabi frequency of $\Omega/2\pi = 1.0$ MHz that dictates the overall energy scale and the blockade radius $R_b$. We specify the dimensionless ratio $R_b / a$ to set the atom separation $a$, and thereby the Rydberg coupling and the Rydberg blockade spacing. We consider a nearest-neighbor Rydberg blockade $R_b / a \in (1,2)$, using a constant global detuning field $\Delta_\mathrm{glob}$ to tune the system to the $Z_2$ ground state phase region~\cite{Bernien:2017,Keesling:2019,Bluvstein:2021}. For $\Delta_\mathrm{glob} \gg \Omega$ and $\Delta_{\mathrm{loc},j} = 0$, the ground state energy eigenvalue is two-fold degenerate since $n_s$ is even; the two degenerate eigenvectors are the antiferromagnetic $Z_2$ states with a 2-site period and opposite order, which are adiabatically connected to the product states $\ket{10 \hdots 10}$ and $\ket{01 \hdots 01}$. In order to break this degeneracy, we introduce the local detuning field with a staggered pattern
\begin{equation}
\Delta_{\mathrm{loc},j} = (-1)^j \Delta_{\mathrm{loc}} \, .
\label{eq:local_detuning}
\end{equation}
We thus obtain false and true vacuum states separated by energy density $\mathcal{E} \propto \Delta_{\mathrm{loc}}$, with each of the vacua dominated by one of the opposite $Z_2$ product states. The detuning waveforms are then specified by two dimensionless parameters:
\begin{align}
\alpha &= \Delta_\mathrm{glob} / \Omega , \label{eq:alpha} \\
\beta &= \Delta_\mathrm{loc}/\Delta_\mathrm{glob} . \label{eq:beta}
\end{align}
where we take $\alpha > 0$. To explore the decay and annealing regimes, we make particular choices for the initial state, $R_b / a$, $\alpha$, and $\beta$. \Cref{fig:overview} schematically shows the choices for the initial state and $\beta$, as well as the resulting dynamics, all of which will be further elucidated in Sections \ref{sec:decay} and \ref{sec:annealing}.

The staggered field produces an interaction potential between domain walls, commonly referred to as a confining potential, because of its analogue in high-energy physics. Domain walls separate regions of opposite antiferromagnetic order; in the site basis, they arise as adjacent sites with occupations of $00$ or $11$, and introduce energy tradeoffs in the the global detuning, local detuning, and nearest-neighbor Rydberg interaction. A domain, or bubble, is a region bounded by two domain walls. In the staggered field, the interaction potential is confining for false vacuum domains and anticonfining for true vacuum domains. 

At a broad level, the state evolution imprints signatures in magnetic order parameters (OP). Neutral atoms in the blockade regime are in the antiferromagnetic phase. A global quantity that captures the antiferromagnetic character is the Neel OP $N = \langle \hat{N} \rangle$, defined by the operator
\begin{equation}
\hat{N} = \frac{2}{n_s} \sum_j (-1)^j \left( \frac{1}{2} \hat{I}_j - \hat{n}_j \right) = \frac{1}{n_s} \sum_j (-1)^j \hat{\sigma}_{z,j} \, .
\label{eq:neel}
\end{equation}
$N = 1$ for the $Z_2$ product state $\ket{10 \hdots 10}$, $N = -1$ for the $Z_2$ product state $\ket{01 \hdots 01}$, and $N = 0$ for the two states $\ket{0 \hdots 0}$ and $\ket{1 \hdots 1}$, for example. 
At a more detailed level, the state evolution produces bubbles of true vacuum. The density of an antiferromagnetic $k$-bubble $\sigma_k = \langle \hat{\Sigma}_k \rangle$ can be determined from the operator $\hat{\Sigma}_k$, whose form depends on the bubble size $k$. The operators for some values of $k$ are
\begin{align}
\hat{\Sigma}_1 &= \frac{1}{n_s - 2} \sum_{j=1}^{n_s - 2} \left[ ( \hat{n}_j \hat{n}_{j+1} \hat{n}_{j+2} ) + \mathrm{rev} \right] \, , \label{eq:density_of_1_bubbles} \\
\hat{\Sigma}_2 &= \frac{1}{n_s - 3} \sum_{j=1}^{n_s - 3} \left[ ( \hat{n}_j \hat{n}_{j+1} \hat{g}_{j+2} \hat{g}_{j+3} ) + \mathrm{rev} \right] \, , \label{eq:density_of_2_bubbles} \\
\hat{\Sigma}_{n_s} &= ( \hat{n}_1 \hat{g}_2 \hat{n}_3 \hat{g}_4 \hdots \hat{n}_{n_s - 1} \hat{g}_{n_s} ) \, , \label{eq:density_of_ns_bubbles}
\end{align}
where $\hat{g}_j = \hat{I}_j - \hat{n}_j = \outerprod{g_j}{g_j}$, the terms in parentheses have the pattern of one type of $k$-bubble, and the term ``rev'' is obtained by making the reversal $\hat{n}_j \leftrightarrow \hat{g}_j$ for all $j$ to obtain the pattern of the opposite type of $k$-bubble.
\section{Decay}
\label{sec:decay}

False vacuum decay occurs as a tunneling process via bubble nucleation induced by quantum fluctuations~\cite{Coleman:1977a,Coleman:1977b,Devoto:2022}. Above a certain critical size, a true vacuum bubble is driven to expand, since the bulk energy decrease can compensate for the surface energy increase~\cite{Coleman:1977a,Coleman:1977b,Devoto:2022}; such a supercritical bubble will rapidly expand throughout the state, whereas a subcritical bubble cannot. In the finite temperature case, nucleation is driven by both quantum and thermal fluctuations, and the late-time state will be a thermal state rather than the true vacuum. The decay process has interesting applications in many scenarios, including early-Universe dynamics in cosmology~\cite{Kobzarev:1974cp,Coleman:1977a,Coleman:1977b,Callan:1977}, soliton dynamics in high-energy physics~\cite{Stone:1976,Stone:1977,Voloshin:1985id}, and non-equilibrium dynamics in condensed matter systems~\cite{Rutkevich:1999,Lagnese:2021,Zenesini:2024}. 

In this section, we study false vacuum dynamics in the decay regime (\Cref{fig:overview_decay}). Our efforts have several novel features. We investigate the decay process with antiferromagnetic 1D neutral atom systems, and incorporate long-range Rydberg tails. These systems support confinement, spatially inhomogeneous field configurations, and high levels of control, and thus have all the key features to induce decay by nucleation~\cite{Henriet:2020,aquila2023quera}. Our studies are explicitly devised to accommodate system sizes, coherence times, and waveform constraints achievable by near-term neutral atom experiments.

\subsection{Decay Rate}
\label{subsec:decay:rate}

The identifiable signature of tunneling by nucleation is the characteristic scaling of the decay rate. We first review earlier derivations of the decay rate scaling in quantum field theory and spin chains. We then determine the form of the decay rate scaling in our neutral atom setup, in terms of the confinement parameter and the energy gap of the Hamiltonian. 

\subsubsection{Quantum Field Theory}
\label{subsubsec:decay:rate:qft}

In quantum field theory, pioneering efforts to study false vacuum decay examined a scalar field Lagrangian $\mathcal{L} = \frac{1}{2} (\partial_\mu \phi)^2 - V(\phi)$~\cite{Kobzarev:1974cp,Stone:1976,Stone:1977,Coleman:1977a,Coleman:1977b,Callan:1977,Voloshin:1985id,Lee:1986}. 
In the semiclassical limit where $\hbar$ is small, and thus the field is treated as a classical particle and the tunneling process is treated quantum mechanically, the decay rate, i.e. the tunneling probability per unit time, can be written as $\gamma = A e^{-B}$, with $B = S_E [\phi_b]$ for $V(\phi_{FV}) = 0$, where the false vacuum is labeled $\phi_{FV}$ and the ``bounce'' solution $\phi_b$ is a stationary point of the Euclidean action $S_E$~\cite{Kobzarev:1974cp,Stone:1977,Coleman:1977a,Coleman:1977b,Callan:1977,Lee:1986}. In the ``thin-wall'' limit where the critical bubble size is large compared to the bubble thickness, and thus the energy density difference $\mathcal{E}$ between the false and true vacua is small, the decay rate takes a simplified form since the rigorous upper and lower bounds on the decay exponent converge~\cite{Brown:2018}. In the ``thin-wall'' limit, the form of the decay rate can be obtained in the semiclassical limit~\cite{Kobzarev:1974cp,Stone:1977,Coleman:1977a,Coleman:1977b,Callan:1977,Lee:1986} or, in the more general case, using effective action methods~\cite{Stone:1976,Voloshin:1985id}; in $1+1$ spacetime dimensions and in certain potentials, it has the form $\gamma \propto \mathcal{E} e^{-\lambda \mathcal{E}^{-1}}$~\cite{Stone:1977,Voloshin:1985id}, which becomes exponentially small for small $\mathcal{E}$.

The tunneling process in quantum field theory is different from the process in single-particle nonrelativistic quantum mechanics~\cite{Lee:1986,Devoto:2022}. In the latter, the tunneling depends directly on the potential only~\cite{Lee:1986,Devoto:2022}. In the former, the tunneling depends more elaborately on the system properties, including both the states and the potential. The states are field configurations. In $(3+1)$ dimensions, the relevant potential for tunneling is the functional $U[\phi] = \int d^3 x [ \frac{1}{2} (\nabla \phi)^2 + V(\phi) ]$ rather than the field potential $V(\phi)$~\cite{Lee:1986,Devoto:2022}. The functional $U[\phi]$ not only contains an extra gradient term from the states, but also an integration over all space. It thereby prohibits a direct transition between two spatially homogeneous field configurations, since the integral produces an infinite energy barrier. Tunneling thus proceeds through nucleation, in which the immediate post-tunneling state is not spatially homogeneous. The tunneling process in quantum field theory exhibits distinct behavior; for instance, it can even occur without a potential barrier~\cite{Lee:1986}.

\subsubsection{Spin Chains}
\label{subsubsec:decay:rate:spins}

False vacuum decay has also been examined in quantum many-body systems with confinement. In particular, it has been studied in the ferromagnetic Ising spin model~\cite{Rutkevich:1999,Lagnese:2021}, described by the Hamiltonian
\begin{equation}
H_I = - \sum_{i=1}^N ( \hat{\sigma}_{z,i} \hat{\sigma}_{z,i+1} + h_x \hat{\sigma}_{x,i} + h_z \hat{\sigma}_{z,i} ) \, ,
\label{eq:hamiltonian_ising}
\end{equation}
where $h_x$ and $h_z$ are the transverse and longitudinal fields. For $h_z = 0$, the ground state is two-fold degenerate; domain walls propagate freely with dispersion relation $\omega(\theta) = 2 ( 1 - 2 h_x \cos\theta + h_x^2 )^{1/2}$ where $\theta$ is the quasimomentum. In the further case $0 < |h_x| < 1$ and zero temperature limit, the system is ferromagnetic and the two ground states $\ket{0_{\pm}}$ have magnetizations $\langle \hat{\sigma}_{z,i} \rangle = \pm \mathcal{M}$ where $\mathcal{M} = (1 - h_x^2)^{1/8}$. For $h_z \neq 0$, the ground state degeneracy is broken; domain walls experience a linear interaction potential, or confinement potential. The resulting true and false vacuum states have energy density difference, or confinement energy density, proportional to $\mathcal{E} \propto h_z \mathcal{M}$. 

The decay rate was derived rigorously in the thin-wall limit $|\mathcal{E}| \ll 1$, or equivalently $|h_z| \ll 1/\mathcal{M}$, using the integrability of the model at $h_z = 0$, yielding~\cite{Rutkevich:1999}
\begin{equation}
\gamma \propto N |h_z| \mathcal{M} g(h_z) e^{-|f(\theta_0)| / ( |h_z| \mathcal{M} ) } \, ,
\label{eq:decay_rate_Ising_spins}
\end{equation}
where $g(h_z)$ is a slowly-varying function of $h_z$ under common assumptions, $f(\theta) = \int_0^\theta \omega(\alpha) d\alpha$, and $\theta_0 = i |\ln{h_x}|$ is the imaginary root of $\omega(\theta)$. This expression was derived in a 2D limit that holds for a 1D system. Notably, the term $f(\theta_0)$ in the numerator is specific to the exactly solvable zero confinement system. Recent numerical simulations with quantum quenches have corroborated the exponential term~\cite{Lagnese:2021}.
This expression is analogous to the one from $(1+1)$-dimensional quantum field theory, if we substitute the confinement energy density $\mathcal{E} \propto h_z \mathcal{M}$. However, the specific form depends on the detailed microphysics of the system.

The form of the decay rate in these systems can be understood intuitively with a simplified picture of the tunneling process~\cite{Lagnese:2021}. A critical bubble of true vacuum has size $\ell \propto \mathcal{E}^{-1}$, a result that can be obtained by comparing the classical energy of the false vacuum state to that of a state with a true vacuum bubble. For small $\mathcal{E}$, a small bubble is produced by nucleation, and then grows incrementally into a critical bubble. This process is nonperturbative, and thus the matrix element for exciting critical bubbles exhibits exponential scaling in $\mathcal{E}^{-1}$; the decay rate follows. At finite temperature, the critical bubbles will thermalize and the late-time state will be a thermal state, not the true vacuum~\cite{Lagnese:2021}. 

The form of the decay rate scaling in \Cref{eq:decay_rate_Ising_spins} extends beyond Ising spin chains to other condensed matter systems in the perturbative regime, after setting $\mathcal{E}$ to the relevant confinement energy density. Notably, it has been corroborated numerically in the XXZ ladder~\cite{Lagnese:2021} and experimentally with ferromagnetic superfluids formed from atomic Bose-Einstein condensates~\cite{Zenesini:2024}.

\subsubsection{Neutral Atom Setup}
\label{subsubsec:decay:rate:neutral_atoms}

We investigate the decay rate scaling for our neutral atom setup. We impose confinement using the staggered local detuning field in \Cref{eq:local_detuning}, which yields a confinement energy density $\mathcal{E} \propto \Delta_\mathrm{loc}$. The size of a critical bubble can be discerned using a classical picture. If the false vacuum state has energy $\epsilon_0$, then, ignoring the contributions from $V_2$, the energy of a state with a single true vacuum bubble of size $n$ is $\epsilon_{[n]} \approx \epsilon_0 + \Delta_\mathrm{glob} - n \Delta_\mathrm{loc}$, yielding a critical bubble size $\ell \sim \Delta_\mathrm{glob} / \Delta_\mathrm{loc}$. We are interested in the ``thin-wall'' limit $|\mathcal{E}| \ll \Delta_\mathrm{glob}$, or equivalently $| \Delta_\mathrm{loc} | \ll \Delta_\mathrm{glob}$, where the size of a critical bubble is much larger than the thickness of the bubble walls. The Hamiltonian we use is not integrable, so there is no closed-form expression for the zero confinement dispersion relation. However, we can estimate its energy scale as follows. The bandwidth $\sigma$ of the first excitation band is proportional to the domain wall two-step hopping energy $T_{n,n+2} \sim \Omega^2 / \Delta_\mathrm{glob} + \Omega^2 / V_1$. For the parameter regime that we examine in \Cref{subsec:decay:numerical}, the zero confinement energy gap $\delta$ is much larger than the bandwidth $\sigma$. We can thus estimate the zero confinement dispersion relation and the analogous numerator term as $f \propto \omega \approx \delta + O(\sigma)$. In the ``thin-wall'' limit $|\mathcal{E}| \ll \Delta_\mathrm{glob}$, we thus expect a decay rate scaling
\begin{equation}
\gamma \propto e^{-\lambda \delta | \beta |^{-1}} \, ,
\label{eq:decay_rate_neutral_atoms}
\end{equation}
where $\lambda$ collects the remaining terms in the exponent. 

\subsection{Numerical Results}
\label{subsec:decay:numerical}

\begin{figure}
\includegraphics[width=0.48\textwidth]{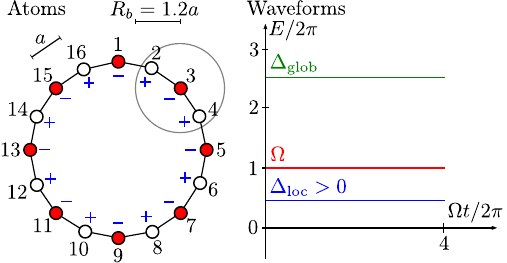}
\caption{The numerical setup for false vacuum decay. The left section shows the atom configuration. The grey circle shows the blockade radius. The initial state is set to the $Z_2$ product state $\ket{\psi(0)} = \ket{10 \hdots 10}$, represented by the color pattern for the atoms. The right section shows the waveforms, all held constant. The local detuning parameter is set to a modest constant positive value $\beta > 0$, shown by the blue $+/-$ symbols on the left and waveform on the right, so that the initial $Z_2$ product state is close to the false vacuum state.}
\label{fig:setup_decay}
\end{figure}

Our numerical setup is shown in \Cref{fig:setup_decay}. We set the initial state to the $Z_2$ product state $\ket{\psi(0)} = \ket{10 \hdots 10}$, shown in the atom chain in \Cref{fig:setup_decay}. We set the local detuning parameter to a modest constant positive value $\beta > 0$, shown by the blue pattern in \Cref{fig:setup_decay}, such that the initial state is close to the false vacuum state. 

The parameter requirements for the decay regime are given by three broad sets of conditions: 
(1) $V_2 \ll \Omega < \Delta_\mathrm{glob} \sim V_1$, so that the system is modestly in the $Z_2$ phase and the Rydberg blockade does not heavily suppress nucleation; 
(2) $\Delta_\mathrm{glob} < | \Delta_\mathrm{loc} | n_s$, so that critical bubbles of size $\ell \sim \Delta_\mathrm{glob} / \Delta_\mathrm{loc} < n_s$ can be nucleated; and
(3) $| \Delta_\mathrm{loc} | \ll \Delta_\mathrm{glob}$, so that the system is in the thin-wall regime, and thus nucleation proceeds by the incremental growth of a domain, and does not produce two or more domains. 
We find a wide parameter range that satisfies these requirements and produces decay. It is centered around the following values: a small nearest-neighbor blockade radius $R_b / a \approx 1.2$ that produces a modest nearest-neighbor interaction $V_1$ comparable to $\Omega$ and $\Delta_\mathrm{glob}$; a modest global detuning parameter $\alpha \in [2.5, 3.5]$; and a modest constant positive local detuning parameter $\beta \in [0.1, 0.5] > 0$ that permits a perturbative description and nucleation of supercritical bubbles, since $\Delta_\mathrm{loc} \ll \Delta_\mathrm{glob} < | \Delta_\mathrm{loc} | n_s$.

To measure the decay rate, we require an observable that takes two disparate values for the false and true vacua, and smoothly transitions for intermediate states. The N\'{e}el OP $N$ in \Cref{eq:neel} satisfies the required properties. 
It is also simple to compute numerically.

\begin{figure*}
\subfloat[$R_b / a = 1.20$, $\alpha = 2.5$]{\label{fig:neel_vs_t}\includegraphics[width=0.312\textwidth]{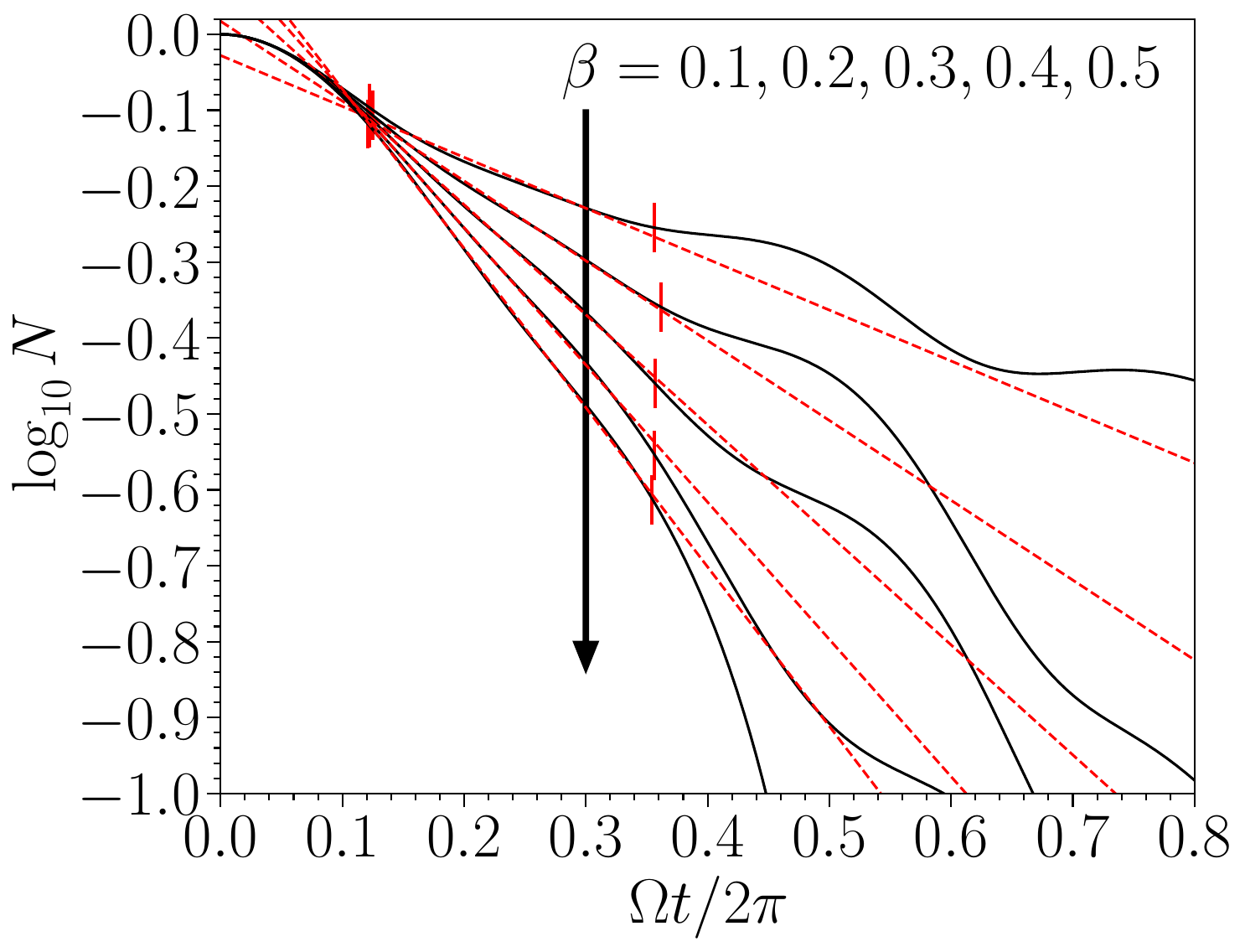}}\hfill
\subfloat[$R_b / a = 1.20$]{\label{fig:gamma_vs_betaInv}\includegraphics[width=0.306\textwidth]{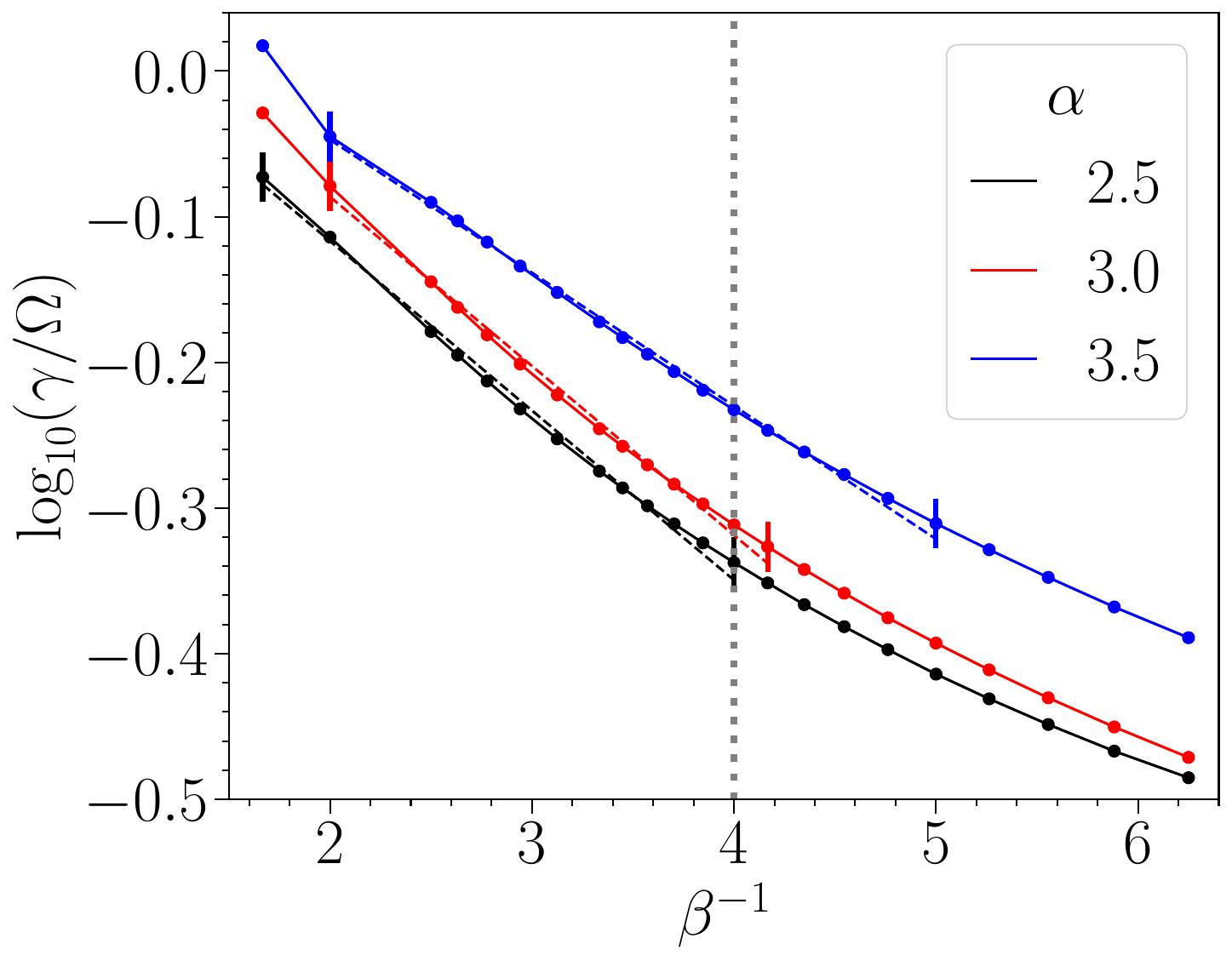}}\hfill
\subfloat[$\beta = 0.25$]{\label{fig:gamma_vs_E20}\includegraphics[width=0.36\textwidth]{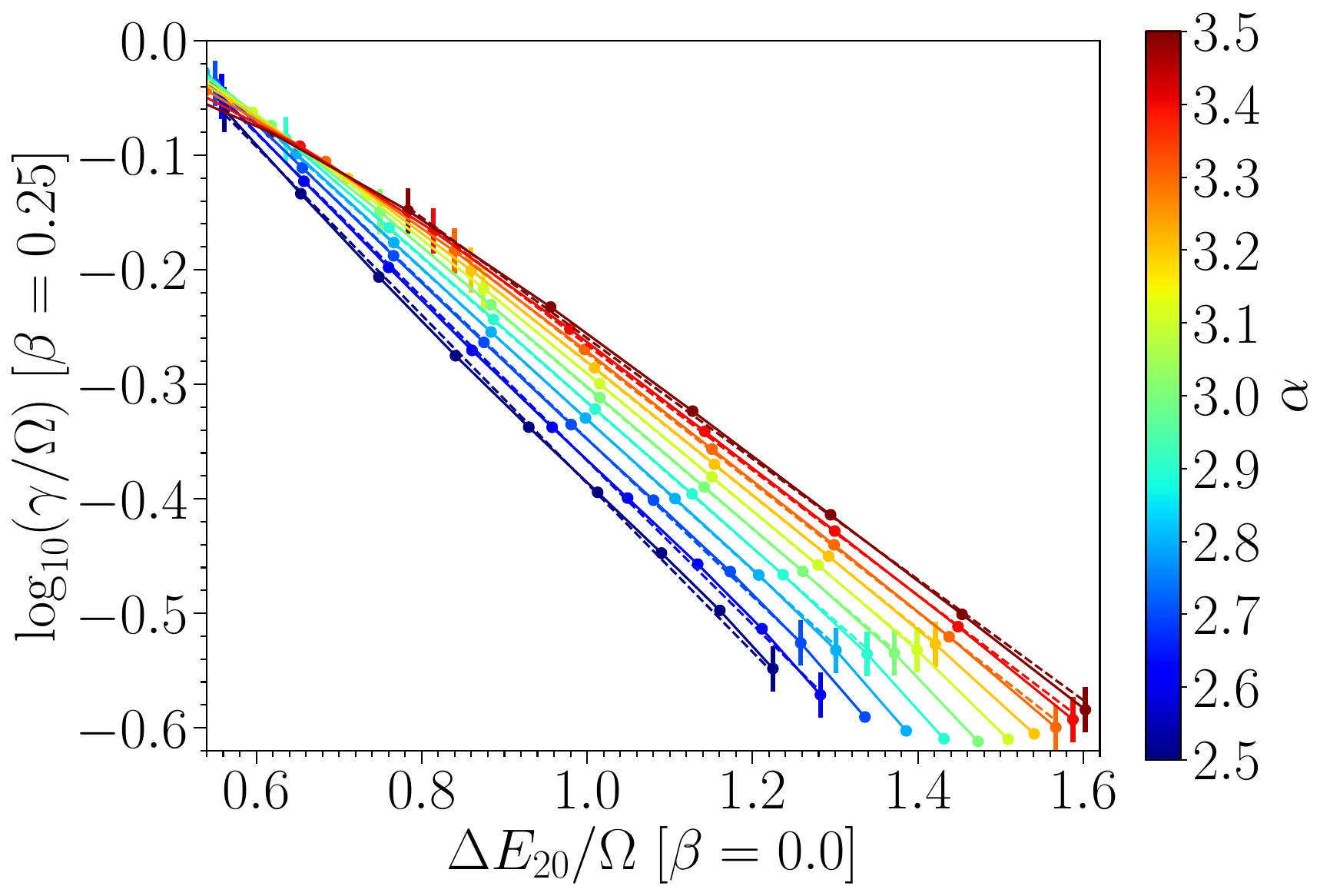}}
\caption{The decay properties of the N\'{e}el OP. In each panel, we perform a least-squares fit to an exponential function; the small vertical bars show the fitting regions and the dashed curves show the fit. (a) The (log) N\'{e}el OP $N$ vs the (dimensionless) time $\Omega t / 2\pi$. The curves correspond to $\beta = 0.1, 0.2, 0.3, 0.4, 0.5$ from top to bottom. The curves exhibit the exponential decay given in \Cref{eq:N_of_t}, characteristic of resonant tunneling as described in \Cref{subsec:decay:rate}. (b) The (dimensionless) decay rate $\gamma/\Omega$ vs the inverse local detuning parameter $\beta^{-1}$. The vertical grey dotted line shows the value $\beta^{-1} = 4.0$ ($\beta = 0.25$). The decay rate exhibits the characteristic scaling with the confinement parameter predicted for false vacuum decay, given in \Cref{eq:decay_rate_neutral_atoms}. (c) The (dimensionless) decay rate $\gamma/\Omega$ [for $\beta = 0.25$] vs the energy gap $\Delta E_{20} = E_2 - E_0$ [for $\beta = 0$], the latter obtained by changing $R_b / a$. It exhibits the derived gap scaling, given in \Cref{eq:decay_rate_neutral_atoms}.}
\label{fig:neel_decay}
\end{figure*}

\begin{figure*}
\subfloat[Ground state phase diagram for $\beta = 0$]{\label{fig:phase_diagrams:gs}\includegraphics[width=0.33\textwidth]{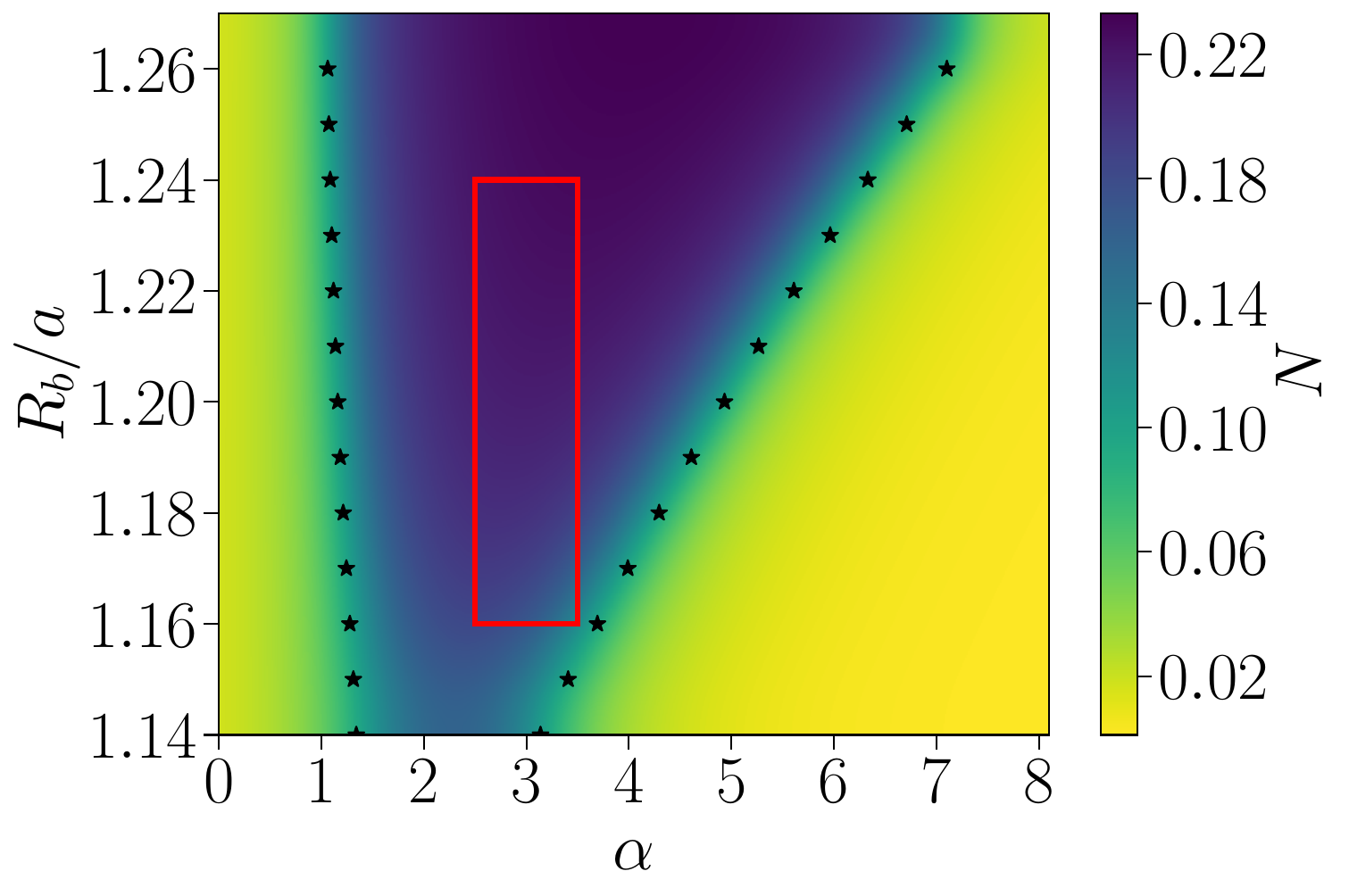}}\hfill
\subfloat[Ground state phase diagram for $\beta = 0$]{\label{fig:phase_diagrams:gs_restricted_range}\includegraphics[width=0.33\textwidth]{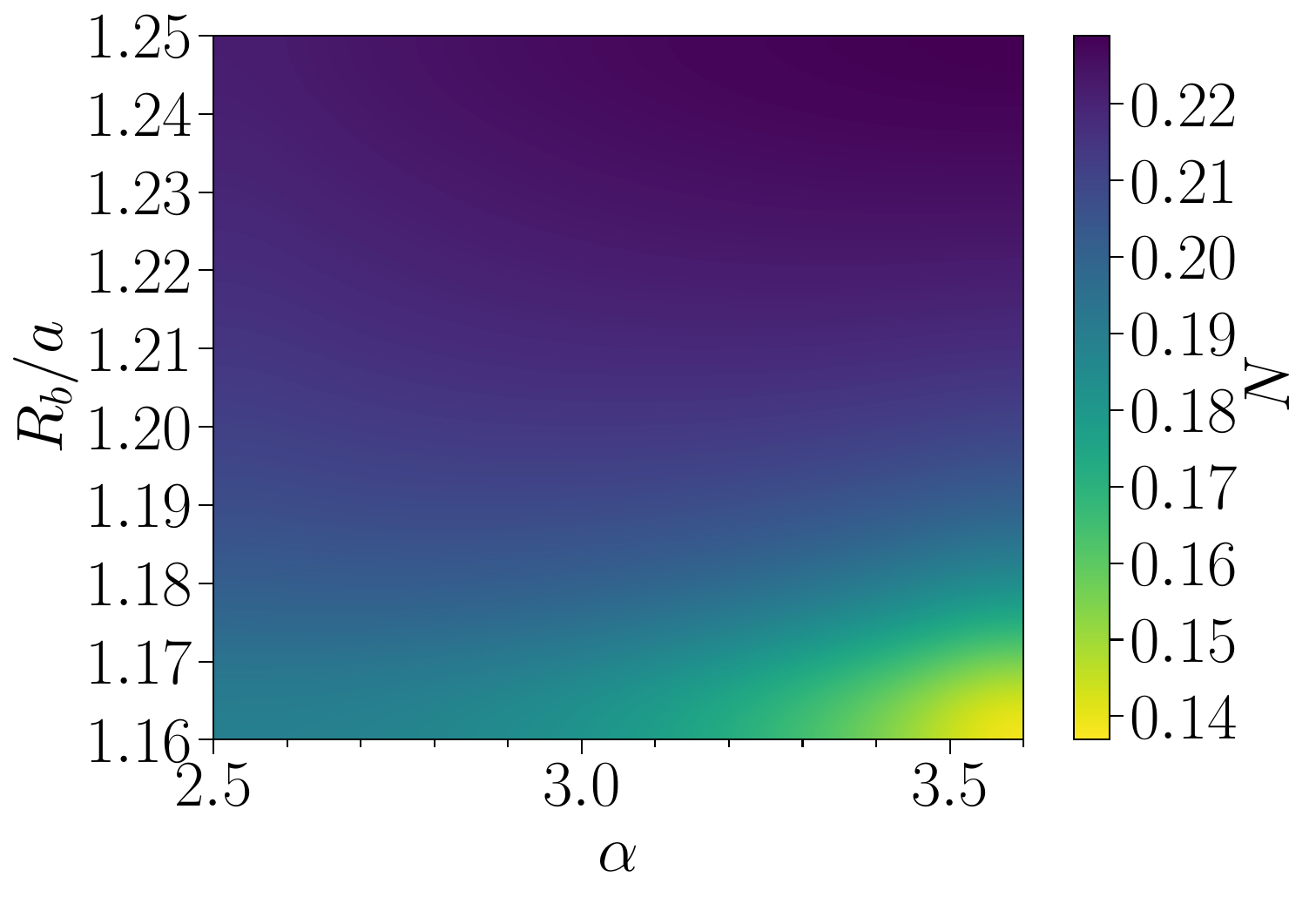}}\hfill
\subfloat[Decay rate $\gamma / \Omega$ for $\beta = 0.25$]{\label{fig:phase_diagrams:gamma}\includegraphics[width=0.32\textwidth]{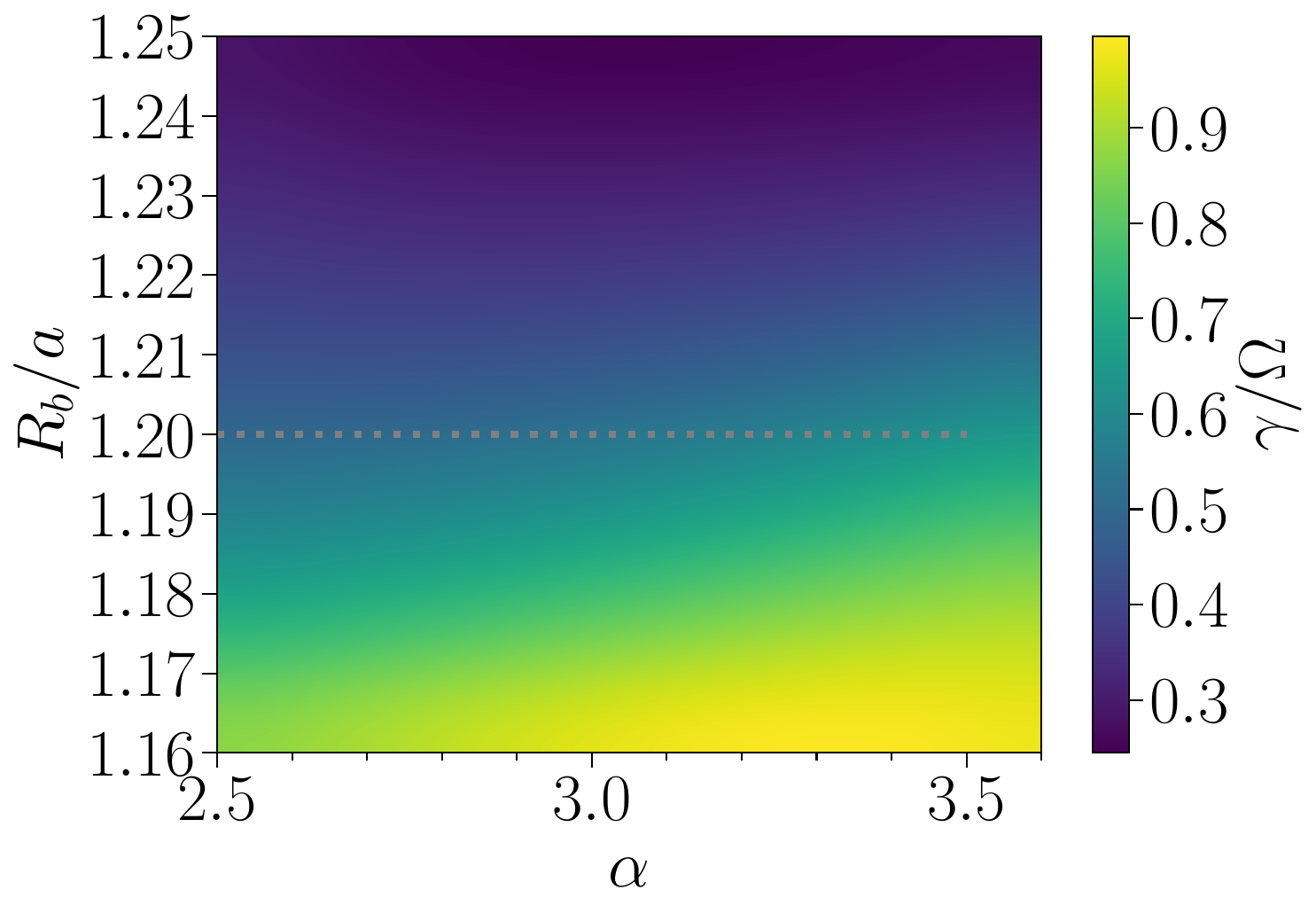}}
\caption{The phase properties in the decay regime. (a) The ground state phase diagram for $\beta = 0$, presented in terms of the N\'{e}el OP $N$ of the two-point correlation function (TPCF) $g_{ij}$, defined analogously to \Cref{eq:neel}. The black stars show the phase transition points, calculated from the inflection points of the curves of constant $R_b / a$. The red rectangle shows the boundaries of \ref{fig:phase_diagrams:gs_restricted_range} and \ref{fig:phase_diagrams:gamma}. (b) The same as \ref{fig:phase_diagrams:gs} for the range in the red rectangle. (c) The (dimensionless) decay rate $\gamma / \Omega$ for $\beta = 0.25$. The horizontal grey dotted line shows the value $R_b / a = 1.20$. The decay rate qualitatively follows the behavior of the N\'{e}el OP, which is a consequence of the underlying gap scaling.}
\label{fig:phase_diagrams}
\end{figure*}

\Cref{fig:neel_vs_t} shows a representative sample of the early-time evolution of the N\'{e}el OP $N$, which undergoes exponential decay within the rough time interval $\Omega t / 2\pi \in [0.1, 0.4]$, in agreement with the theoretical prediction from tunneling. We perform a least-squares fit to the exponential 
\begin{equation}
N = A e^{-\gamma t}
\label{eq:N_of_t}
\end{equation}
to obtain the decay rate $\gamma$ and the amplitude $A$. To determine the fitting interval, we compute a smoothed N\'{e}el OP by applying a Savitzky-Golay filter with polynomial order $n=3$ and calculate the $(10 - 90) \%$ amplitude range of the smoothed curve in the time interval $\Omega t / 2\pi \in [0.1, 0.4]$. The curves experience an initial transient at the early times $\Omega t / 2\pi \lesssim 0.1$, during which the false vacuum is stationary and subcritical bubbles nucleate and incrementally grow~\cite{Lagnese:2021}. The curves exhibit oscillations at later times $\Omega t / 2\pi \gtrsim 0.4$, due to the evolution and thermalization of the post-decay state. 

The decay rate exhibits exponential scaling with the inverse local detuning (\Cref{fig:gamma_vs_betaInv}), within an interval that minimally contains $\beta^{-1} \in [2.5, 4.0]$. We thus observe the phenomenological predicted scaling with the confinement parameter determined in \Cref{eq:decay_rate_neutral_atoms}. We can compare this interval with our expected parameter condition for the decay regime. The lower limit of the interval gives $\Delta_\mathrm{loc} = 0.4 \Delta_\mathrm{glob}$ and the upper limit of the interval gives $\Delta_\mathrm{loc} n_s = 4.0 \Delta_\mathrm{glob}$, which satisfy our condition $\Delta_\mathrm{loc} \ll \Delta_\mathrm{glob} < | \Delta_\mathrm{loc} | n_s$. The interval for the confinement scaling depends on the system size. We set the fitting interval by inspection and perform a least-squares fit to the exponential $\gamma = b e^{-p \beta^{-1}}$, where $b$ and $p$ are fitting parameters. Though not shown, the curves become more closely spaced at larger $R_b / a$. 

We examine in detail the properties of the decay rate for a specific local detuning value; we select $\beta = 0.25$ ($\beta^{-1} = 4.0$). In \Cref{fig:gamma_vs_betaInv}, this value is shown by the vertical grey dotted line, falling towards the end of the fitting range of the confinement parameter, though we note that this figure only presents results for $R_b / a = 1.20$. In our subsequent discussion, we consider a range of values of $R_b / a$.

The decay rate for the system with $\beta = 0.25$ exhibits exponential scaling with the energy gap for system with $\beta = 0$, as shown in \Cref{fig:gamma_vs_E20}. For zero confinement $\beta = 0$, the energy gap is $\Delta E_{20} = E_2 - E_0$, where $E_2$ is the second excited state and $E_0$ is the ground state. The different values of $\Delta E_{20}$ correspond to different values of $R_b / a$. As discussed in \Cref{subsubsec:decay:rate:neutral_atoms}, the zero confinement dispersion relation is approximately equal to the energy gap $\delta = \Delta E_{20}$ up to first order corrections in the bandwidth $\sigma$, and thus these results corroborate our theoretically expected gap scaling in \Cref{eq:decay_rate_neutral_atoms}. In contrast to the confinement scaling, the interval for the gap scaling should remain valid if the domain wall hopping energy is much less than the potential energy, and thus be independent of system size. We set the fitting interval by inspection and perform a least-squares fit to the exponential $\gamma = k e^{-q (\Delta E_{20} / \Omega)}$, where $q$ and $k$ are fitting parameters. Though not presented, the parameter $q$ appears to show a linear scaling with $\alpha$: it fits the linear function $q = u (\alpha_0 - \alpha)$ for fitting parameters that satisfy $u < 0$ and $\alpha_0 > \alpha$. We cannot discern the cause of this subleading linear dependence in terms of nucleation theory.

The decay rate for $\beta = 0.25$ can be further compared to the Hamiltonian ground state structure for $\beta = 0$ by considering the phase properties (\Cref{fig:phase_diagrams}). \Cref{fig:phase_diagrams:gs} shows the ground state phase diagram for the equivalent Hamiltonian with zero local detuning $\beta = 0$. The ground state phase is presented by the N\'{e}el OP of the two-point correlation function (TPCF) $g_{ij}$ of the ground state, defined analogously to the N\'{e}el OP of the Rydberg densities $n_j$ in \Cref{eq:neel}. The blue central region indicates the antiferromagnetic phase. The yellow outer regions indicate the disordered phase. The two boundaries between them indicate phase transitions; the black stars show several transition points, determined by examining curves of constant $R_b / a$ and calculating the points with vanishing second derivative. \Cref{fig:phase_diagrams:gs_restricted_range} simply shows the ground state phase diagram over a smaller axis range, that bounded by the red rectangle in \Cref{fig:phase_diagrams:gs}. 

\Cref{fig:phase_diagrams:gamma} shows the decay rate for $\beta = 0.25$. It provides a complementary picture to \Cref{fig:gamma_vs_betaInv}, which presented results for $R_b / a = 1.20$; the horizontal grey dotted line in the former corresponds to the vertical grey dotted line in the latter. The axis range is the false vacuum decay region that we explored, and sets the axis range chosen for \Cref{fig:phase_diagrams:gs_restricted_range} and the red rectangle in \Cref{fig:phase_diagrams:gs}. The false vacuum decay region lies entirely within the antiferromagnetic phase of the ground state in \Cref{fig:phase_diagrams:gs}; this solidifies our decay results and suggests that false vacuum decay is a dynamical transition separate from any ground state phase transitions. The figure thus presents a decay phase diagram within the parameter region that we explored. The main features of this decay phase diagram arise from the scaling of the decay rate with the energy gap. The gap closes at the ground state phase boundaries in \Cref{fig:phase_diagrams:gs}. The gap also grows with the ground state N\'{e}el OP shown in \Cref{fig:phase_diagrams:gs_restricted_range}, and thus the corresponding decay rate decreases in \Cref{fig:phase_diagrams:gamma}. Therefore, though our initial state is near the false vacuum state and not the true vacuum, the decay rate still qualitatively probes the features of the ground state phase diagram. The energy gap of the system is hard to access in neutral atom experiments, since it precludes preparation of excited states. However, the decay rate of the false vacuum appears to be a possible route for qualitative comparisons of gap sizes. Furthermore, when the absolute gap size is unnecessary, such as for critical scalings, the decay rate of the false vacuum could even be used for a quantitative analysis.
\section{Annealing}
\label{sec:annealing}

In lieu of a quench to the decay regime, a slow ramp of the confining field can also nucleate quantized bubbles. Sinha et al.~\cite{Sinha:2021} recently proposed and studied this process in 1D Ising spin chains, with the Hamiltonian given by \Cref{eq:hamiltonian_ising}. In particular, they initialized the state in the false vacuum, tuned a linear ramp of the confining field $h_z$, and observed sequential nucleation of $n$-bubbles at the resonance points $h_z = 2/n$. They further found that the resonances occur at Landau-Zener (LZ) avoided crossings. 

In this section, we study nucleation dynamics of the false vacuum in 1D neutral atom chains under a linear ramp of the confining field $\Delta_\mathrm{loc}$. However, instead of the ramp direction used in Ref.~\cite{Sinha:2021}, we use the opposite direction to obtain a reversed energy landscape for false and true vacuum states. Though one might expect this ramp to adiabatically tune the false vacuum of the initial system to the true vacuum of the final system, we instead uncover a ramp that anneals the false vacuum from the initial system to the final system, with intermediate nucleation events corresponding to condensation and decondensation points. Our procedure demonstrates an adiabatic passage through excited states, and can be generalized to other systems. For instance, it can be considered in the cosmological context. If the energy splitting of the false and true vacua slowly changes sign, then the Universe will remain in a false vacuum state. However, our results show that even this slow annealing case can produce nucleation events that are potentially observable.

\subsection{Numerical Results}
\label{subsec:annealing:numerical}

\begin{figure}
\includegraphics[width=0.48\textwidth]{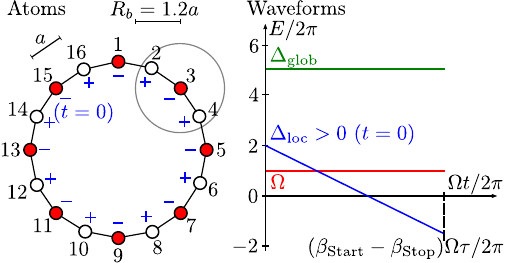}
\caption{The numerical setup for false vacuum annealing. The left section shows the atom configuration. The grey circle shows the blockade radius. The initial state is set to the $Z_2$ product state $\ket{\psi(0)} = \ket{10 \hdots 10}$, represented by the color pattern for the atoms. The local detuning parameter $\beta = \Delta_\mathrm{loc} / \Delta_\mathrm{glob}$ is set to a linear ramp $\beta(t) = \beta_\mathrm{Start} - t/\tau$, with $\beta_\mathrm{Start} > 0$ and $\beta_\mathrm{Stop} < 0$, shown at time $t=0$ by the blue $+/-$ symbols on the left and label on the right.}
\label{fig:setup_annealing}
\end{figure}

\begin{figure*}
\subfloat[N\'{e}el order parameter $N$]{\label{fig:N_vs_chi}\includegraphics[width=0.495\textwidth]{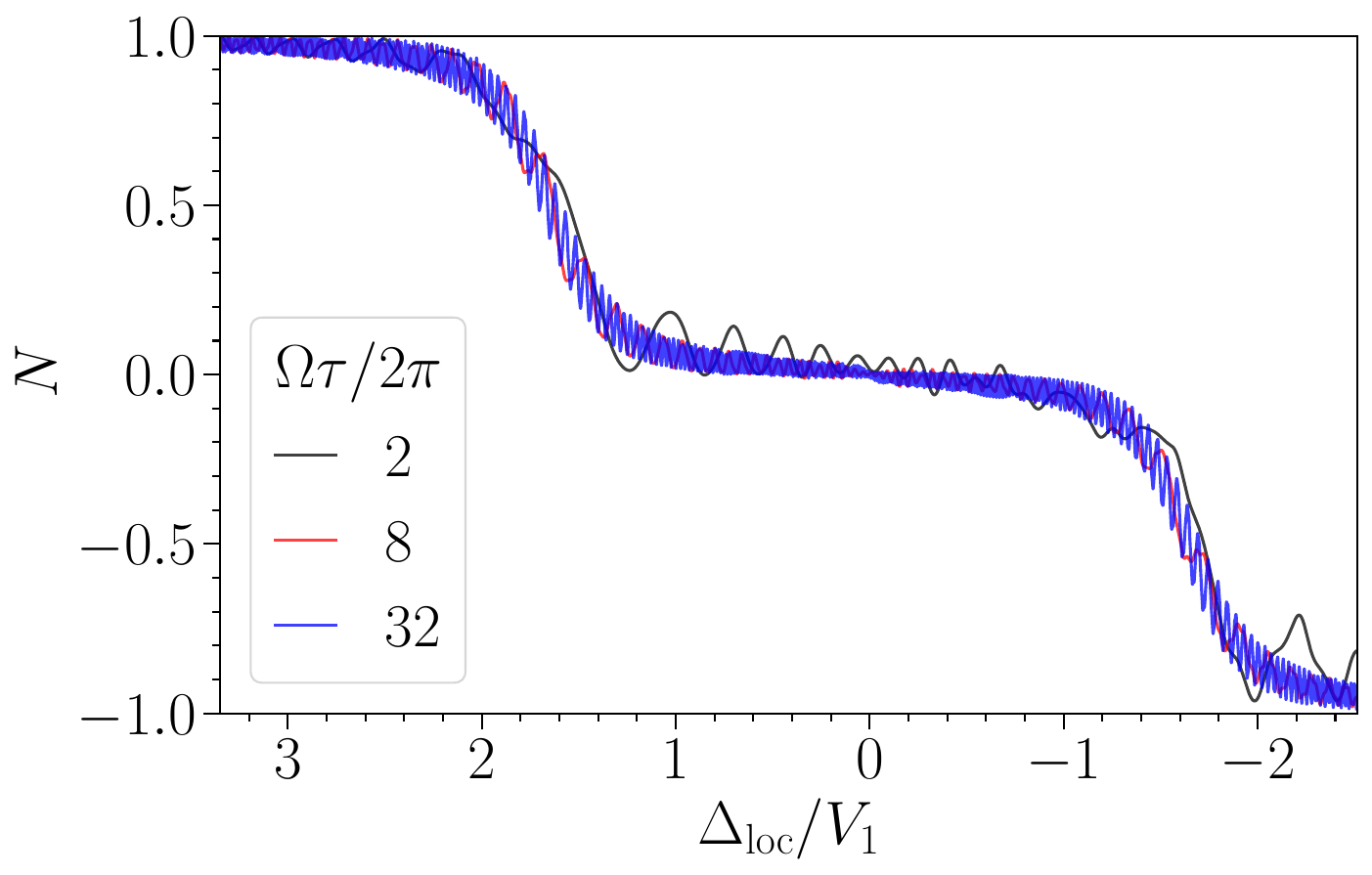}}\hfill
\subfloat[Density of 1-bubbles $\sigma_1$]{\label{fig:sigma1_vs_chi}\includegraphics[width=0.48\textwidth]{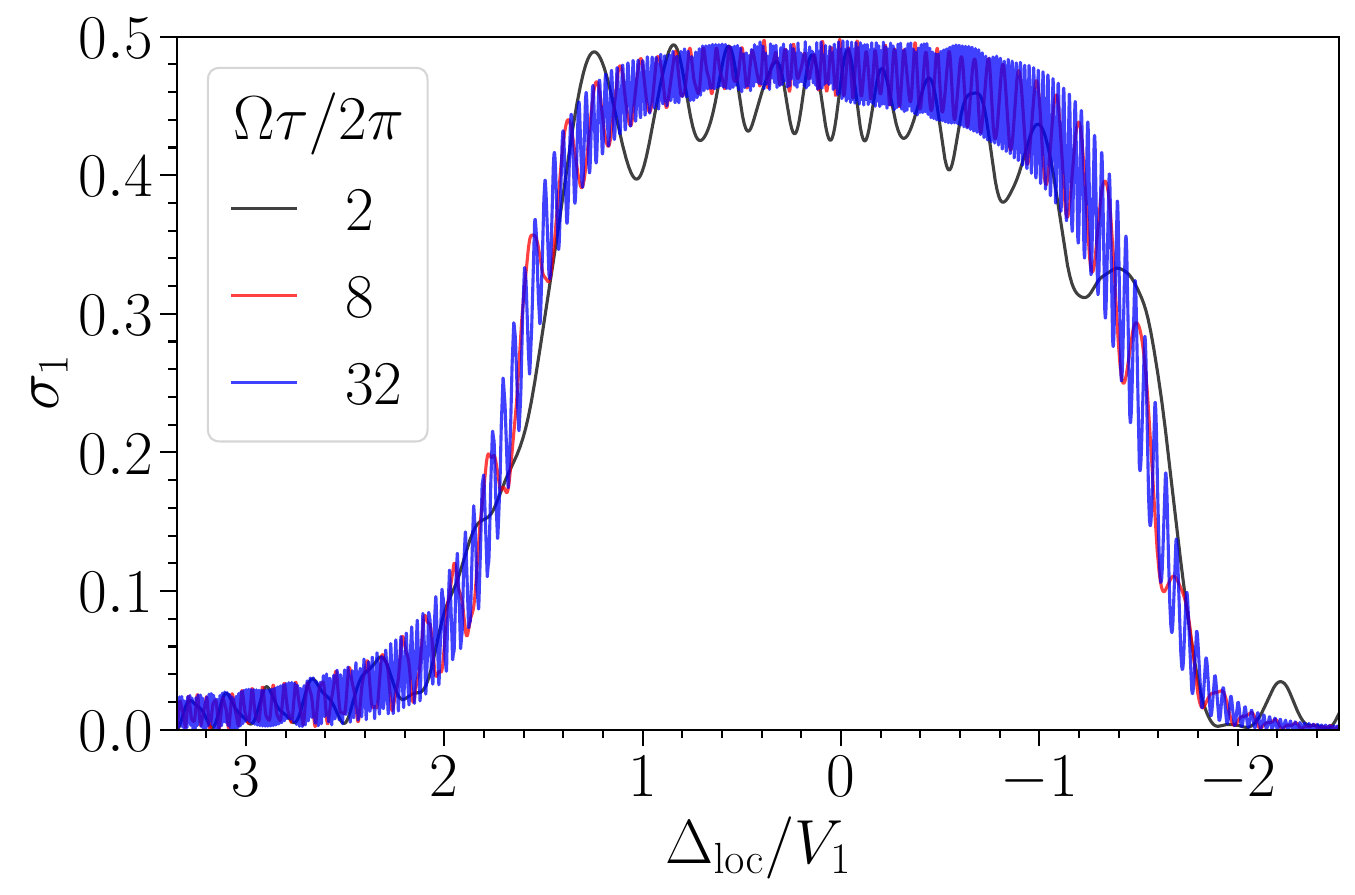}}\\
\subfloat[Density of 2-bubbles $\sigma_2$]{\label{fig:sigma2_vs_chi}\includegraphics[width=0.495\textwidth]{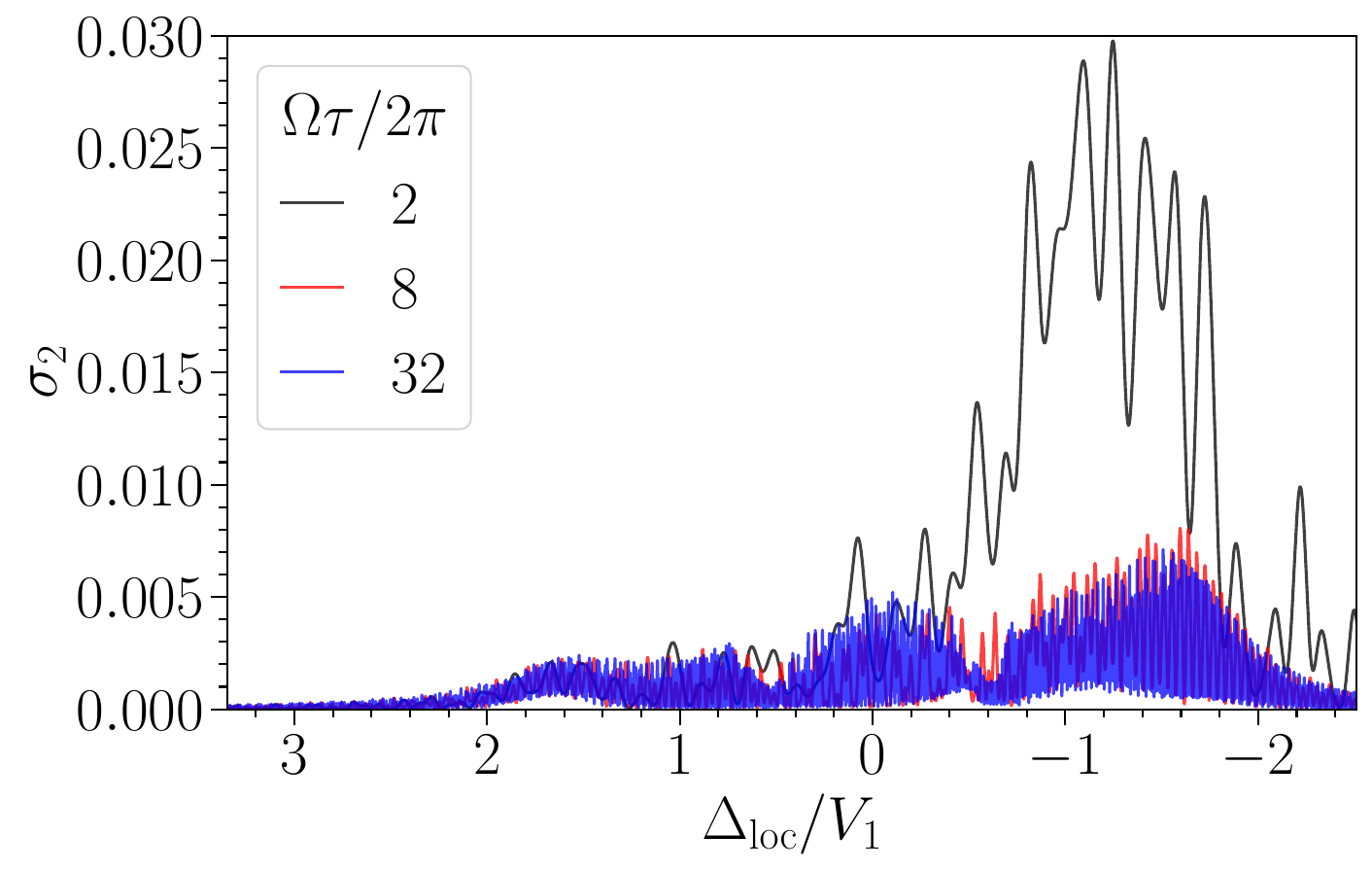}}\hfill
\subfloat[Density of $n_s$-bubbles $\sigma_{n_s}$]{\label{fig:sigmans_vs_chi}\includegraphics[width=0.48\textwidth]{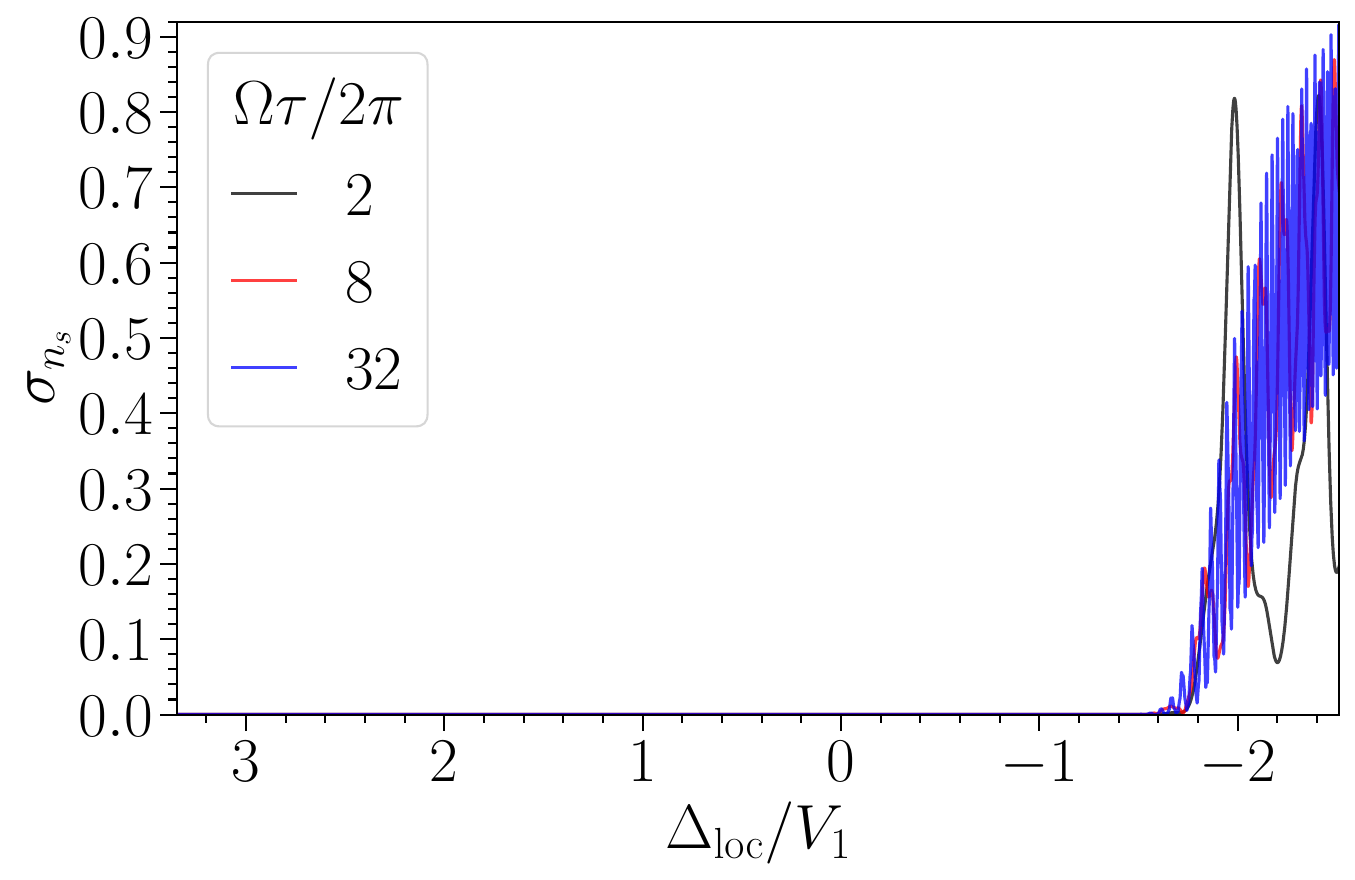}}\\
\caption{Nucleation dynamics during the annealing procedure, for $\alpha = 5.0$ and $R_b / a = 1.2$. In each panel, the local detuning on the $x$-axis is normalized to the nearest-neighbor interaction strength $V_1$, and the different curves correspond to different ramp times $\tau$. (a) The N\'{e}el OP $N$ vs the (dimensionless) local detuning $\Delta_\mathrm{loc} / V_1$. It exhibits plateaus and cliffs, suggesting nucleation events. (b) The density of $1$-bubbles $\sigma_1$, from \Cref{eq:density_of_1_bubbles}. (c) The density of $2$-bubbles $\sigma_2$, from \Cref{eq:density_of_2_bubbles}. (d) The density of $n_s$-bubbles $\sigma_{n_s}$, from \Cref{eq:density_of_ns_bubbles}. The densities of the $k$-bubbles show that the first cliff is a condensation point that approximately produces the state $\ket{0 \hdots 0}$, the all zero state, and the second cliff is a decondensation point that approximately produces the state $\ket{01 \hdots 01}$, the false vacuum of the final Hamiltonian.}
\label{fig:bubble_nucleation}
\end{figure*}

\begin{figure*}
\subfloat[$H_r$ eigenvalues]{\label{fig:two_atom_model:eigenvalues}\includegraphics[width=0.48\textwidth]{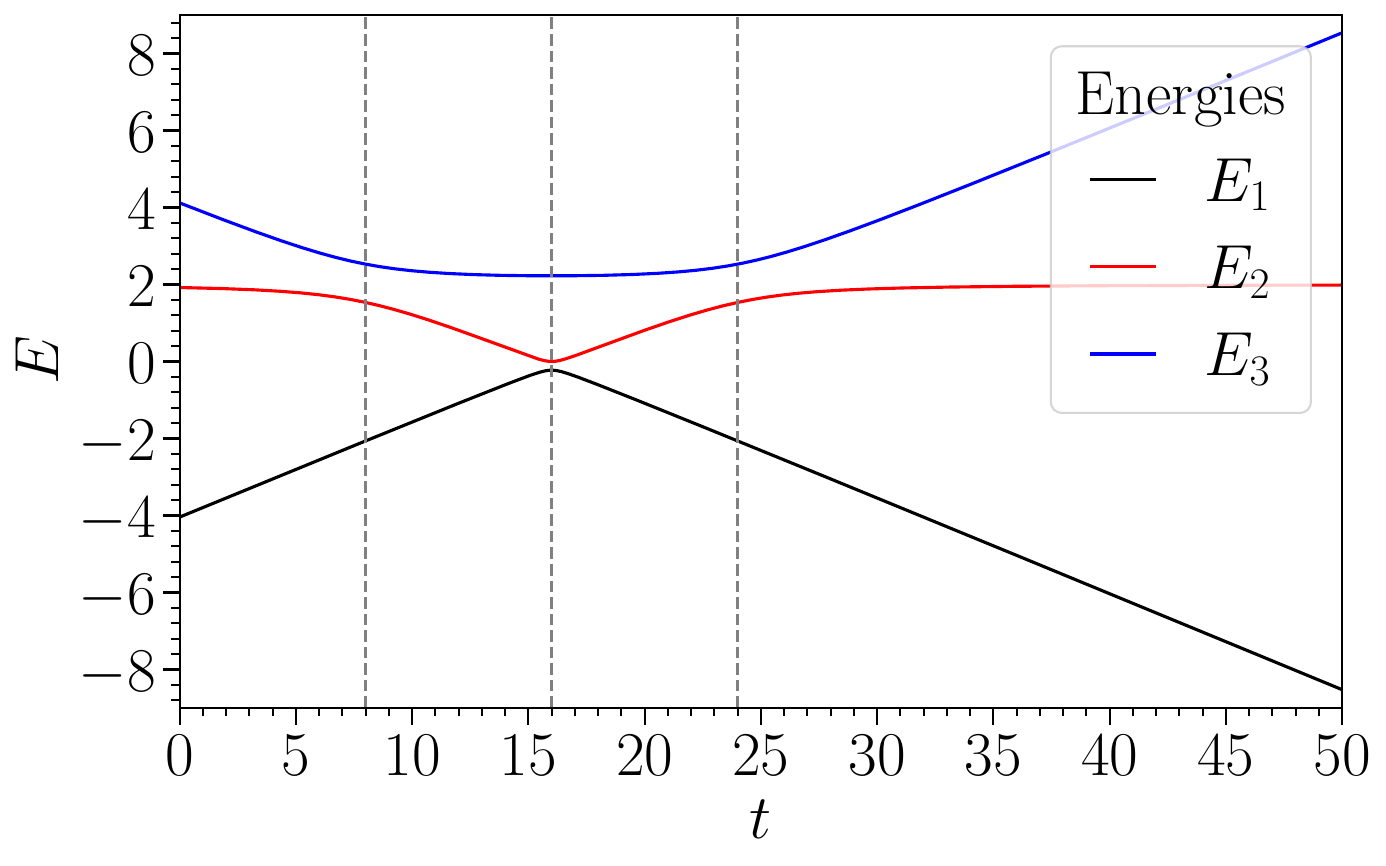}}\hfill
\subfloat[Coefficients]{\label{fig:two_atom_model:coefficients}\includegraphics[width=0.485\textwidth]{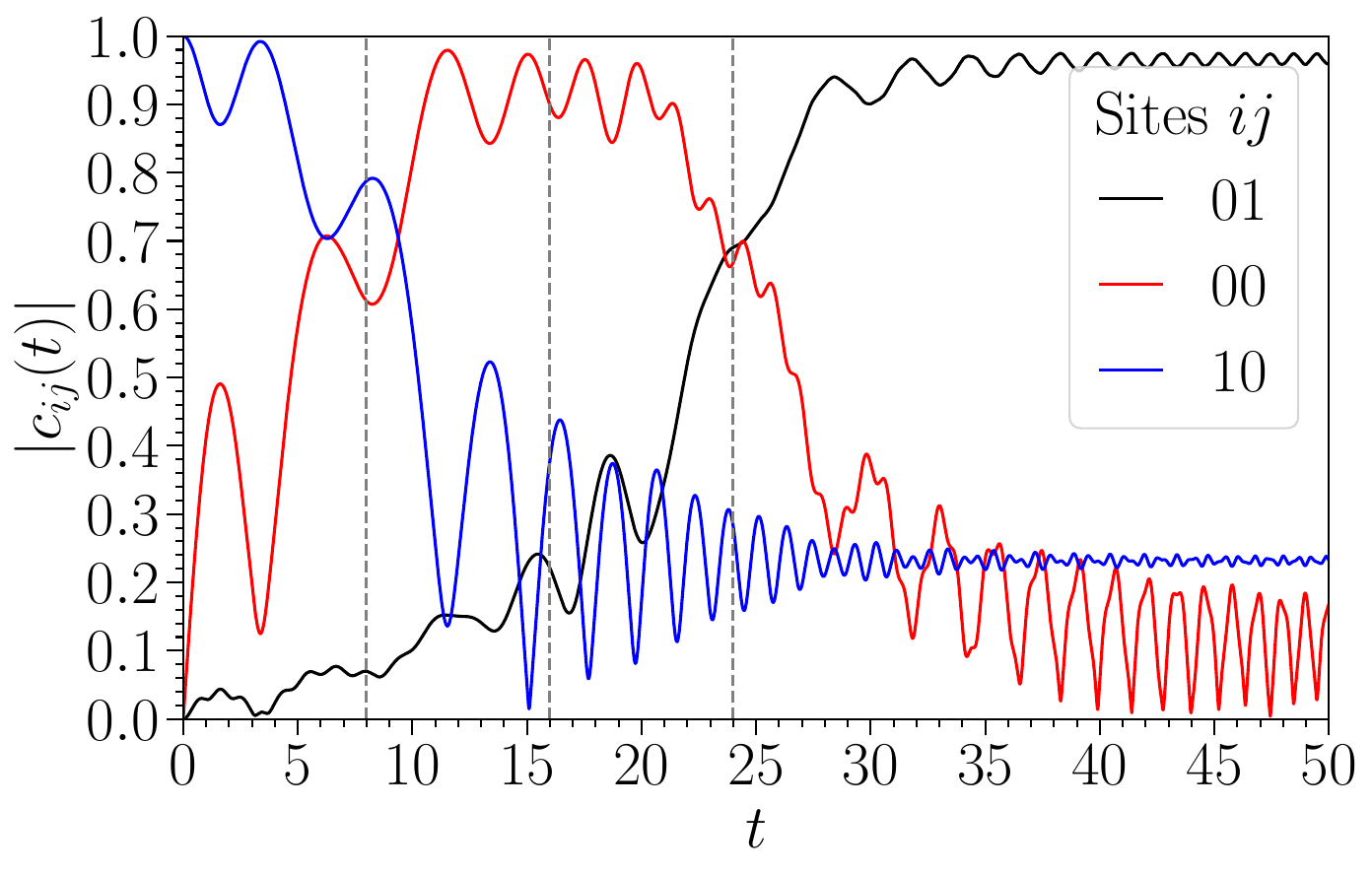}}
\caption{The evolution of the eigenvalues and state in the two-atom model. We set $\Omega = 1$. (a) The energy eigenvalues of $H_r$ vs time $t$. The eigenvalue $E_3 (t)$ corresponds to the false vacuum state $\ket{\phi_3 (t)}$. (b) The coefficients $c_{ij}(t)$ of the state $\ket{\psi(t)}$ in the site basis vs time $t$. The initial state is the $Z_2$ product state $\ket{\psi(0)} = \ket{10}$. It has overlap probability $p = | \langle \phi_3 (0) | \psi(0) \rangle |^2 > 0.97$ with the false vacuum state at $t=0$. In both panels, the middle vertical grey dashed line shows the time $t = \beta_\mathrm{Start} \tau$ (i.e. $\Delta_\mathrm{loc} = 0$), and the left and right vertical grey dashed lines show the times $t = (\beta_\mathrm{Start} \pm 1) \tau$ (i.e. $\Delta_\mathrm{loc} = \pm \Delta_\mathrm{glob}$), which are the locations of the Landau-Zener (LZ) avoided crossings for $E_3 (t)$. Since we adiabatically tune the local detuning with the linear ramp, the evolution of $\ket{\psi(t)}$ is largely determined by the adiabatic evolution of $\ket{\phi_3 (t)}$ and $E_3 (t)$. The evolution in this two-atom model explains the nucleation behavior of the many-atom system shown in \Cref{fig:bubble_nucleation}.}
\label{fig:two_atom_model}
\end{figure*}

We outline our numerical setup in \Cref{fig:setup_annealing}. We set the initial state to the $Z_2$ product state, $\ket{\psi(0)} = \ket{10\hdots10}$. We set the local detuning parameter to the linear ramp 
\begin{equation}
\beta(t) = \beta_\mathrm{Start} - t/\tau \, ,
\end{equation}
where $\beta_\mathrm{Start}$ is the start amplitude and $\tau$ is the ramp timescale. A larger value of $\tau$ implies a slower ramp. The local detuning can be substituted for the time since the two are linearly related by the ramp, and can be normalized to the nearest-neighbor interaction strength $V_1$ to quantify its relative strength. 

We use the following parameter values: a small nearest-neighbor blockade spacing $R_b / a = 1.2$, a large global detuning parameter $\alpha = 5.0$, large local detuning start and stop amplitudes $\beta_\mathrm{Start} = 2.0$ and $\beta_\mathrm{Stop} = -1.5$, and local detuning ramp times $\tau \in [2, 32]$.

We aim to study the nucleation of $k$-bubbles during the annealing procedure. We thus characterize the evolved state by the N\'{e}el OP $N$ given in \Cref{eq:neel} and the density of $k$-bubbles $\sigma_k$ given in Equations \eqref{eq:density_of_1_bubbles} - \eqref{eq:density_of_ns_bubbles}. 

The N\'{e}el OP $N$ evolves from $N = 1$ to $N = -1$ during the procedure, through plateaus punctuated by rapid cliffs (\Cref{fig:N_vs_chi}). The curves evolve similarly for different values of the ramp time $\tau$, though curves with lower ramp times exhibit stronger oscillations, as expected. The procedure thus appears stable for antiferromagnetic systems. 

The densities of $k$-bubbles reveal several features that clarify the nucleation dynamics and the behavior of the N\'{e}el OP. Figures \ref{fig:sigma1_vs_chi} to \ref{fig:sigmans_vs_chi} show the densities of the $k$-bubbles defined in Equations \eqref{eq:density_of_1_bubbles} to \eqref{eq:density_of_ns_bubbles}. The first plateau $\Delta_\mathrm{loc} / V_1 \in [2.0, 3.5]$ exhibits gradual nucleation of $1$-bubbles, and a small signal of $2$-bubbles. The first cliff $\Delta_\mathrm{loc} / V_1 \in [1.6, 2.0]$ coincides with the rapid nucleation of 1-bubbles to its maximum density $\sigma_1 \approx 0.5$, indicating that it is a condensation point that approximately produces the state $\ket{0 \hdots 0}$. The first cliff also sees the low-density production and destruction of $2$-bubbles. During the second plateau $\Delta_\mathrm{loc} / V_1 \in [-1.2, 1.5]$, the density of 1-bubbles remains near its maximum value $\sigma_1 \approx 0.5$, but slightly declines as the local detuning increases. This slight decline is precipitated by the gradual low-density nucleation of $2$-bubbles (and $3$-bubbles more prominently, though not shown). The second cliff $\Delta_\mathrm{loc} / V_1 \in [-2.0, -1.2]$ and subsequent plateau $\Delta_\mathrm{loc} / V_1 \in [-2.5, -2.0]$ coincide with rapid nucleation of a $n_s$-bubble near its maximum value $\sigma_{n_s} \approx 1$, indicating that is a decondensation point that approximately produces the state $\ket{01 \hdots 01}$. The second cliff also sees the low-density production and destruction of $2$-bubbles (and $3$-bubbles, though not shown). The final state is then near the false vacuum of the final Hamiltonian, dominated by the opposite $Z_2$ product state $\ket{01 \hdots 01}$. 

\subsection{Two-Atom Model}
\label{subsec:annealing:model}

To explain the nucleation dynamics, we examine a 2-atom chain. In short, we find that the nucleation events correspond to Landau-Zener (LZ) avoided crossings.

We examine the 2-atom chain using only the nearest-neighbor blockade subspace, which yields a restricted Hilbert space spanned by the states $\{ \ket{00}, \ket{01}, \ket{10} \}$ and the restricted Hamiltonian
\begin{equation}
\begin{split}
H_r = & \Delta_\mathrm{glob} \outerprod{00}{00} - \Delta_\mathrm{loc} (t) ( \outerprod{01}{01} - \outerprod{10}{10} ) \\
& + \frac{\Omega}{2} ( \outerprod{01}{00} + \outerprod{10}{00} + \mathrm{h.c.} ) \, ,
\end{split}
\end{equation}
where we added an overall energy shift of $\Delta_\mathrm{glob}$ to simplify terms. The restricted Hamiltonian describes a gapped three-level system that is simple enough to study analytically and yet rich enough to capture the features observed in the many-atom case. 

We set the initial state to the $Z_2$ product state $\ket{\psi(0)} = \ket{10}$. We set the Rabi frequency to $\Omega = 1$, the global detuning to $\Delta_\mathrm{glob} = 2.0$, and the local detuning ramp parameters to $\beta_\mathrm{Start} = 2.0$ and $\tau = 8.0$. 

We can understand the important nucleation times by considering the evolution of the three energy eigenvalues $E_i (t)$ of $H_r$ during the procedure (\Cref{fig:two_atom_model:eigenvalues}). Each eigenvalue is symmetric about $t = \beta_\mathrm{Start} \tau$ (or $\Delta_\mathrm{loc} = 0$). The eigenvalues $E_1 (t)$ and $E_2 (t)$ exhibit a LZ avoided crossing at this point. The eigenvalues $E_2 (t)$ and $E_3 (t)$ exhibit LZ avoided crossings at $t = (\beta_\mathrm{Start} \pm 1) \tau$ (or $\Delta_\mathrm{loc} = \pm \Delta_\mathrm{glob}$).

The false vacuum state is the energy eigenstate $\ket{\phi_3 (t)}$ corresponding to $E_3 (t)$. At time $t=0$, it has overlap probability $p = | \langle \phi_3 (0) | \psi(0) \rangle |^2 > 0.97$ with the initial state $\ket{\psi(0)} = \ket{10}$. We adiabatically tune the local detuning with the linear ramp, and thus the evolution of the state $\ket{\psi(t)}$ is largely determined by the adiabatic evolution of $\ket{\phi_3 (t)}$, which is punctuated by the LZ avoided crossings of $E_3 (t)$. \Cref{fig:two_atom_model:coefficients} shows the evolution of the state. The procedure nucleates a 1-bubble at the first LZ avoided crossing $t = (\beta_\mathrm{Start} - 1) \tau$ (or $\Delta_\mathrm{loc} = -\Delta_\mathrm{glob}$) to produce the state $\ket{\psi(t)} \approx \ket{00}$. It then nucleates a 2-bubble at the second LZ avoided crossing $t = (\beta_\mathrm{Start} + 1) \tau$ (or $\Delta_\mathrm{loc} = +\Delta_\mathrm{glob}$) to produce the state $\ket{\psi(t)} \approx \ket{01}$, which has large overlap probability with the false vacuum of the final Hamiltonian.

The two-atom model thus explains the features of the many-atom case in \Cref{subsec:annealing:numerical}. The initial state $\ket{\psi(0)} = \ket{10 \hdots 10}$ is close to the false vacuum state of the initial Hamiltonian. The false vacuum energy has two main LZ avoided crossings. The first one nucleates 1-bubbles to produce the state $\ket{\psi (t)} \approx \ket{00 \hdots 00}$, and due to the larger system size it also produces and destroys nearby 2-bubbles at low density. The second one nucleates a $n_s$-bubble to produce the state $\ket{\psi (t)} = \ket{01 \hdots 01}$, which is close to the false vacuum state of the final Hamiltonian, and due to the larger system size it also produces and destroys 2- and 3-bubbles at low density.
\section{Experimental Protocols}
\label{sec:experiments}

In this section, we investigate how to extend our numerical studies to explicit experimental protocols on current and near-term neutral atom systems, including analog quantum simulators~\cite{Henriet:2020,aquila2023quera}. The protocols must accommodate hardware constraints. Some current constraints on a prototypical system are~\cite{aquila2023quera}: a maximum simulation time limit $t = 4$ $\mu$s; a minimum atomic spacing $a_\mathrm{min} = 4.0$ $\mu$m, including $a_{\mathrm{min},y} = 4.0$ $\mu$m in the vertical direction, with expected near-term upgrades to $2.0$ $\mu$m; a field of view $(F_x, F_y) = (75.0, 76.0)$ $\mu$m, with expected near-term upgrades to $(75.0,120.0)$ $\mu$m; a Rabi frequency $\Omega \in [0.00, 15.8]$ rad/$\mu$s, with maximum slew rate $250.0$ rad/$\mu$s$^2$; and a global detuning $\Delta_\mathrm{glob} \in [-125.0, 125.0]$ rad/$\mu$s, with maximum slew rate $2500.0$ rad/$\mu$s$^2$. Importantly, constraints on the local detuning field introduce limitations on the confining potential. The most consequential experimental limitation is the number of local detuning channels that an experiment can support, with each channel representing a local detuning term of the form:
\begin{equation}
\Delta^\mathrm{ch}_j (t) = f_j \Delta^\mathrm{ch} (t) \, ,
\label{eq:local_detuning_channel}
\end{equation}
where $f_j$ is a time-independent spatial pattern. Experiments that employ at least two local detuning channels can implement our protocol more directly than those that do not, purely through pulse engineering. In these cases, one local detuning channel can be used for the preparation of the desired initial false vacuum state, and another for the quench. However, each local detuning channel represents a new hardware and control element, introducing significant cost and complexity to experiments. Thus,  we describe in detail protocols for more common near-term experiments that only require a single local detuning channel. While we devise our setups specifically with false vacuum decay and nucleation experiments in mind, we expect them to be applicable to a range of quantum simulation tasks that require local fields that differ in the initial state preparation and quench dynamics phases, notably several lattice gauge theory simulation proposals~\cite{Surace:2020,Surace:2021,Homeier:2023}. 

\subsection{Decay}
\label{subsec:experiments:decay}

\begin{figure*}
\subfloat[Waveform sequences]{\label{fig:protocol_decay:waveforms}\includegraphics[width=0.32\textwidth]{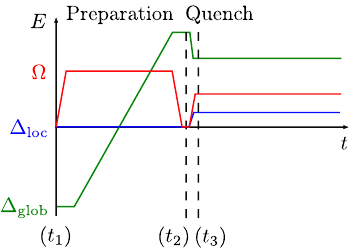}}\hfill
\subfloat[Time $t_1$]{\label{fig:protocol_decay:atom_configuration}\includegraphics[width=0.32\textwidth]{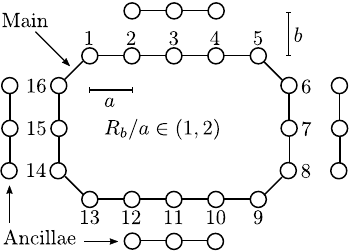}}\hfill
\subfloat[Time $t_3$]{\label{fig:protocol_decay:post_quench_state}\includegraphics[width=0.32\textwidth]{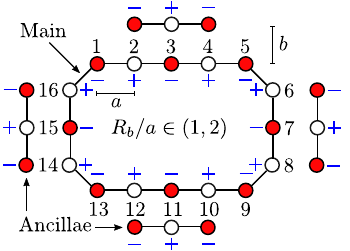}}
\caption{A possible protocol for false vacuum decay experiments on neutral atom systems. The atoms are configured in a main chain with $n_s = 16$ atoms and periodic boundary conditions, and ancilla chains to aid preparation of the $Z_2$ state. (a) The waveforms vs time $t$, with labels on three important times: the initial time $t_1$, quench time $t_2$, and post-quench time $t_3$. (b) The state at the initial time $t_1$. (c) The prepared $Z_2$ state and the applied staggered local detuning field (blue $+/-$) at the post-quench time $t_3$.}
\label{fig:protocol_decay}
\end{figure*}

In \Cref{subsec:decay:numerical}, we examined a numerical setup for the decay process with the following input parameters: $\Omega / 2\pi = 1.0$ MHz ($R_b = 9.76$ $\mu$m); $a \in [8.00, 8.27]$ $\mu$m; $\Delta_\mathrm{glob} / 2\pi \in [2.5, 3.5]$ MHz; and $\Delta_\mathrm{loc} / 2\pi \in [0.625, 1.4]$ MHz. In this input range, the decay process itself occurs on a timescale $t_\mathrm{decay} \approx 0.4 - 0.8$ $\mu$s.

For setups that do not support multi-channel local detuning quenches, we present a potential simulation protocol in \Cref{fig:protocol_decay}. The waveform sequence is shown in \Cref{fig:protocol_decay:waveforms} and consists of two standard pieces: a preparation stage and a quench stage. The preparation stage aims to prepare the desired $Z_2$ state connected to the $Z_2$ product state $\ket{\psi(0)} = \ket{10 \hdots 10}$. The quench stage aims to produce the desired evolution Hamiltonian, and thus involves a quench to obtain the waveforms in our numerical setup. The atom arrangement consists of a main chain and ancilla chains, as shown in \Cref{fig:protocol_decay:atom_configuration}.

The preparation stage uses an adiabatic ramp of the global detuning, as shown in the left part of \Cref{fig:protocol_decay:waveforms}. The main chain has atom separation $a$. It has an even number $n_s = 16$ sites, so in isolation the two $Z_2$ states would be degenerate. The odd-site ancilla chains around the main chain serve to suppress the opposite $Z_2$ state and preferentially produce the desired $Z_2$ state. The ancilla chains have the same atom separation $a$ as the main chain and sit at a distance $b$ from the main chain, where $b \approx a + \epsilon$ for some small $\epsilon$, so that the main-ancilla interaction is weaker than the nearest-neighbor interaction on the main chain. If possible, a preferable distance would be $b \approx R_b + R_\epsilon$, so that the ancilla chain is outside the blockade radius altogether. As an estimate, the distance can be assigned the rough value $b \approx 10.0$ $\mu$m. Informed by the experiments that studied quantum many-body scars~\cite{Bernien:2017,Bluvstein:2021}, we estimate a preparation time of at least $t_\mathrm{prep} \approx 2.0$ $\mu$s to produce a $Z_2$ state with sufficient contrast. 

The quench stage involves a quench of all the waveforms, including a single-sign quench of the local detuning fields, as shown in the right part of \Cref{fig:protocol_decay:waveforms}. The post-quench configuration is shown in \Cref{fig:protocol_decay:post_quench_state}, and includes the prepared $Z_2$ state and the post-quench local detuning pattern. The quench requires large slew rates on the waveforms; irregularities can be minimized by turning off the Rabi drive during the quench. Given the estimated preparation time, we estimate a post-quench time of $t_\mathrm{evolve} \approx 2.0$ $\mu$s, which is longer than the required decay timescale $t_\mathrm{decay} \approx 0.4 - 0.8$ $\mu$s. Together with the preparation time of 2.0 $\mu$s, we are well within the coherence time of current devices~\cite{aquila2023quera}. In the post-quench period, the Rabi frequency $\Omega / 2\pi = 1.0$ MHz implies a maximum local detuning phase accumulation of $3\pi$. To efficiently freeze the evolution of the ancilla atoms in the post-quench period, one can apply a significantly higher staggered local detuning field on the ancilla atoms than on the main chain, perhaps up to $\Omega' / 2\pi = 10$ MHz, while not affecting the coherence of the main chain. If we extend the atom configuration in the vertical direction, the setup will require modified post-quench waveform parameters to produce false vacuum decay. 

The atom configuration for $n_s = 16$ shown in \Cref{fig:protocol_decay:atom_configuration} fits within the field of view $(F_x , F_y ) = (75.0, 76.0)$ $\mu$m. For the largest value $a = 8.27$ $\mu$m and $b \approx 10$ $\mu$m, it has dimensions $d_x = (4+\sqrt{2})a + 2b \approx 64.8$ $\mu$m and $d_y = (2+\sqrt{2})a + 2b \approx 48.2$ $\mu$m. We can also run the protocol for a longer main chain. The atom geometry was chosen so that the horizontal width is maximal and the vertical width can be extended. It must be extended vertically in steps of 4 atoms, required to preserve the pattern on the main chain and the relative locations of the ancilla chains. For an upgraded $F_y = 120$ $\mu$m, the field of view can accommodate up to $n_y = 9$ atoms along the vertical part of the main chain, for a main chain size $n_s = 2 n_x + 2 n_y = 28$. Longer chains should produce the same decay rate scaling, but would offer more space to tune $\Delta_\mathrm{loc}$ relative to $\Delta_\mathrm{glob}$, and thus test the decay rate scaling over a broader relative range. However, the fidelity of initial state preparation decreases as the chain size increases. 

For setups that do support multi-channel local detuning quenches, the protocol is more direct. In the preparation stage, one applies a staggered local detuning field with one pattern to the chain to obtain the desired $Z_2$ state. In the quench stage, one performs a quench by turning on the second local detuning channel to obtain the staggered local detuning field with the opposite pattern.

\subsection{Annealing}
\label{subsec:experiments:annealing}

In \Cref{subsec:annealing:numerical}, we examined a numerical setup for the annealing process with the following input parameters: $\Omega / 2\pi = 1.0$ MHz ($R_b = 9.76$ $\mu$m); $a = 8.13$ $\mu$m; $\Delta_\mathrm{glob} / 2\pi = 5.0$ MHz; $\Delta_\mathrm{loc,Start} / 2\pi = 10.0$ MHz; and $\Delta_\mathrm{loc,Stop} / 2\pi = -7.5$ MHz. 

For setups that do not support multi-channel local detuning fields, the annealing stage can be executed by performing a linear ramp of the appropriate local detuning field, with the starting point being determined by the ancilla chain. This approach introduces additional complications compared to a direct quench approach accessible with multi-channel detuning fields. Notably, the field to freeze the ancilla chains would also need to be ramped slowly, potentially introducing some undesired ancilla dynamics. However, we expect this to be a secondary effect, still permitting exploration of the false vacuum annealing phenomenon. For setups that do support multi-channel local detuning fields, the protocol is straightforward, as in the decay regime. In both cases, a ramp time $\tau$ should be chosen so that the ramp duration $t_\mathrm{ramp} = (\beta_\mathrm{Start} - \beta_\mathrm{Stop}) \tau$ is less than the expected post-quench time $t_\mathrm{evolve} \approx 2$ $\mu$s. For instance, the ramp time $\tau = 0.5$ $\mu$s corresponds to a ramp duration $t_\mathrm{ramp} = 1.75$ $\mu$s. From our numerical results in \Cref{fig:bubble_nucleation} for the nearby value $\tau = 2$ $\mu$s, we expect that the evolution will track the plateaus and cliffs, though it will be highly oscillatory.
\section{Conclusion and Outlook}
\label{sec:conclusion}

In this paper, we studied false vacuum decay and nucleation using neutral atoms with Rydberg excitations. We focused on two dynamical regimes, and investigated setups that are accessible to current and near-term neutral atom experiments. In the decay regime, we uncovered the parameter range to observe the decay process and corroborated the expected decay rate scaling with the confinement field and the gap size. Similarly, our work in the annealing regime found a procedure to transfer the false vacuum of the initial system to the final system using a linear ramp, passing through condensation and decondensation points that nucleate bubbles, producing observable signatures. We concluded by considering experimental implementations for state preparation and quantum quenches on near-term neutral atom hardware.

This work opens several directions for further exploration. The confinement strategy that we used, i.e. a staggered local detuning field, can inform additional confinement strategies in neutral atom systems and others, such as spin systems~\cite{Lake:2010,Kormos:2017,Liu:2019,Tan:2021}. The parameter regime that we uncovered for false vacuum decay and annealing in Sections \ref{subsec:decay:numerical} and \ref{subsec:annealing:numerical} will motivate parameter selection in experiments. The experimental protocols that we suggested in \ref{sec:experiments} can be implemented on near-term neutral atom platforms, including quantum simulators~\cite{Henriet:2020,aquila2023quera}. The protocols executed on real quantum hardware can accommodate much longer 1D chains, and thus enable investigations as a function of system size. The prefactor in the decay rate could be investigated with more careful studies, which would permit the extraction of more subtle system properties~\cite{Voloshin:1985id,Rutkevich:1999}. The decay rate scaling could also be examined outside of the thin-wall limit~\cite{Munster:2000,Dunne:2005}, by tuning the confinement field to larger values, which would quantify higher-order confinement effects. Our experimental atom configuration, which used a main chain with ancilla chains, can be employed in experimental studies of other dynamical phenomena on 1D neutral atom chains, such as the simulation of lattice gauge theories~\cite{Surace:2020,Surace:2021,Homeier:2023}. False vacuum decay experiments with neutral atoms can also be performed on 2D lattices~\cite{Nogrette:2014,Barredo:2016,Labuhn:2016,Scholl:2021,Ebadi:2021}. These 2D experiments will require modified setups; they will also produce different decay rate scalings, similar to their $(2+1)$-dimensional counterparts in quantum field theory~\cite{Voloshin:2004}. It is difficult to run preparatory classical simulations on large 2D lattices~\cite{Daley:2023,Shaw:2024}, and thus experiments are the primary means to investigate metastability phenomena in these systems. 

The initial investigations of false vacuum decay and nucleation began many decades ago~\cite{Coleman:1977a,Coleman:1977b}. Theoretical efforts, and more recent numerical work, have established several testable predictions. The rapid development of quantum hardware has opened, possibly for the first time, an experimental avenue to interrogate these predictions, more efficiently than any classical computer can, with applications ranging from strongly-correlated quantum matter to cosmology.
\section*{Computing Resources}

We ran our simulations on the high-performance computing system Perlmutter at the National Energy Research Scientific Computing Center (NERSC) based at Lawrence Berkeley National Laboratory~\cite{perlmutter2023nersc} and used the Bloqade software package developed by QuEra Computing~\cite{bloqade2023quera}.

\begin{acknowledgments}
This research was supported by the U.S. Department of Energy (DOE) under Contract No. DE-AC02-05CH11231, through the National Energy Research Scientific Computing Center (NERSC), an Office of Science User Facility located at Lawrence Berkeley National Laboratory. R.V.B. was supported by the Office of Science, Office of Advanced Scientific Computing Research (ASCR) Exploratory Research for Extreme-Scale Science. The numerical study was performed on the high-performance computing system Perlmutter, a NERSC resource, using NERSC award DDR-ERCAP0030190.
\end{acknowledgments}


\bibliography{references} 

\end{document}